\documentclass{aa}
\usepackage{natbib}
\usepackage{graphicx} 
\usepackage{txfonts}
\usepackage{placeins}
\usepackage{ifthen}
\usepackage{sidecap}
\bibpunct{(}{)}{;}{a}{}{,}
\title{Kinematic  and structural  analysis of  the \object{Minispiral}  in the
  Galactic  Center from BEAR  spectro-imagery%
  \thanks{Fig.~\ref{fits-vmap} is also available in FITS format at the CDS via
    anonymous    ftp   to    cdsarc.u-strasbg.fr    (130.79.128.5)   or    via
    http://cdsweb.u-strasbg.fr/cgi-bin/qcat?J/A+A/, see
    Sect.~\ref{online-data}.}%
}

\titlerunning{Kinematic and structural analysis of the \object{Minispiral}}

\author{%
  Thibaut Paumard%
    \inst{1}\fnmsep%
    \thanks{Visiting          Astronomer,
    Canada-France-Hawaii Telescope, operated  by the National Research Council
    of Canada, le  Centre National de la Recherche  Scientifique de France and
    the     University      of     Hawaii.}\fnmsep%
   \thanks{Present   address:   Max-Planck-Institut  f\"ur   extraterrestrische
    Physik,  Postfach  1312,   D-85741  Garching,  Germany.}
  \and  Jean-Pierre Maillard\inst{1}\fnmsep$^{\star\star}$
  \and Mark Morris\inst{2}%
}

\institute{%
 Institut d'astrophysique de Paris (CNRS),  98b Blvd Arago, 75014 Paris, France
  \and
 University of California, Los Angeles, Div.  of Astronomy, Dept of Physics and
  Astronomy, Los Angeles, CA 90095-1562, USA%
}

\offprints{T.~Paumard, \email{paumard@mpe.mpg.de}}

\date{Received 19 August 2003 / Accepted 4 May 2004}

\abstract{Integral  field spectroscopy  of the  inner region  of  the Galactic
  Center, over  a field of roughly $40\arcsec\times40\arcsec$  was obtained at
  2.06~$\mu$m  (\ion{He}{i})  and  2.16~$\mu$m  (Br$\gamma$)  using  BEAR,  an
  imaging Fourier Transform Spectrometer, at spectral resolutions respectively
  of  52.9~km~s$^{-1}$  and  21.3~km~s$^{-1}$,  and a  spatial  resolution  of
  $\simeq~0.5\arcsec$.  The analysis of the data was focused on the kinematics
  of  the   gas  flows,  traditionally   called  the  ``\object{Minispiral}'',
  concentrated   in   the   neighborhood    of   the   central   black   hole,
  \object{Sgr~A$^{\star}$}.   From  the  decomposition into  several  velocity
  components (up to  four) of the line profile extracted at  each point of the
  field, velocity  features were identified.   Nine distinguishable structures
  are  described: the  standard \object{Northern  Arm},  \object{Eastern Arm},
  \object{Bar}, \object{Western  Arc}, and five  additional, coherently-moving
  patches of gas.   From this analysis, the \object{Northern  Arm} appears not
  limited, as usually thought, to  the bright, narrow North-South lane seen on
  intensity  images,   but  it  consists   instead  of  a   weak,  continuous,
  triangular-shaped surface, drawn out into a narrow stream in the vicinity of
  \object{Sgr~A$^{\star}$} where  it shows a  strong velocity gradient,  and a
  bright western rim.  The \object{Eastern Arm} is split into three components
  (a \emph{Ribbon} and  a \emph{Tip}, separated by a  cavity, and an elongated
  feature parallel to the Ribbon:  the \emph{Eastern Bridge}).  We also report
  extinction of  some interstellar  structures by other  components, providing
  information on their relative position along the line of sight.  A system of
  Keplerian orbits can be fitted to most of the \object{Northern Arm}, and the
  bright  rim of this  feature can  be interpreted  in terms  of line-of-sight
  orbit crowding caused  by the warping of the flowing  surface at the western
  edge facing  \object{Sgr~A$^{\star}$}.  These results lead to  a new picture
  of the gas structures in \object{Sgr~A West}, in which large-scale gas flows
  and isolated gas  patches coexist in the gravitational  field of the central
  Black Hole.  The question of the  origin of the ionized gas is addressed and
  a   discussion  of   the  lifetime   of   these  features   is  presented.   
  }

\begin{document}

\maketitle
%________________________________________________________________
\section{Introduction}

Within the inner  2~pc of the Galactic Center (GC)  lies the \object{Sgr~A West}
region, dominated  by ionized gas which,  because of high  obscuration along the
line  of sight,  has  been detected  only  at infrared  and  radio wavelengths.  
Infrared fine-structure line emission  of [\ion{Ne}{ii}] at 12.8~$\mu$m has been
used to  map the gas  distribution a number  of times, with  successively higher
spatial    sampling   and    spatial   and    spectral   resolutions,    up   to
$0.6\arcsec\times1.0\arcsec$  sampling, $\simeq  30$~km~s$^{-1}$  and $2\arcsec$
resolution in  the most recent paper \citep{lacy91}.   In parallel, observations
with the Very Large Array (VLA) telescope provided a 6-cm map of the ionized gas
in  the   radio  continuum   at  $1\arcsec$  resolution   \citep{lo83}.   Later,
\citet{roberts93}  observed the Sgr~A  West complex  in the  radio recombination
H92$\alpha$ line at 3.6 cm (8.3~GHz),  also at a resolution of $1\arcsec$.  Much
higher spatial resolution was reached with the VLA at 13~mm, with a beam size of
$0.15\arcsec\times0.10\arcsec$,  in the course  of a  project to  measure proper
motions  of the  bright,  compact  blobs of  ionized  gas \citep{zhao98}.   This
ionized   region  is   surrounded  by   a   torus  of   neutral  material,   the
\object{Circumnuclear Disk} \citep[CND,][]{liszt83,becklin82,guesten87,yusef01}.

Br$\gamma$ at  2.166~$\mu$m has also  been used to  trace the ionized  gas.  The
first  detection  consisted  of  a  grid  of  spectra  around  \object{GCIRS~16}
\citep{geballe91} which could  not give an overview of  the emission morphology. 
The  availability of near-infrared  arrays has  resulted in  many images  of the
Galactic  Center.   However,  the  ionized  gas  can only  be  detected  in  the
near-infrared by spectro-imaging or by  narrow-band imaging of a strong emission
line.  Broad-band images,  for example in the infrared K  band, are dominated by
the stellar content.  A first  attempt at spectro-imagery in Br$\gamma$ was made
by    \citet{wright89}    with    a    Fabry-Perot    system    scanning    over
$\simeq1\,000$~km~s$^{-1}$, at a modest spectral resolution of 90~km~s$^{-1}$ in
a $38\arcsec\times36\arcsec$  field.  The  data cube obtained  in the  same line
with BEAR, an Imaging Fourier Transform Spectrometer on the Canada-France-Hawaii
telescope represents a  significant effort to cover most  of the central ionized
region  with a  much better  spectral resolution  (FWHM  $21.3$~km~s$^{-1}$), at
seeing-limited   resolution.    A   preliminary   analysis  was   presented   by
\citet{morris00}.   Data   from  the  same  instrument  were   obtained  in  the
2.06~$\mu$m   \ion{He}{i}  line,   leading  to   the  first   identification  of
interstellar    Galactic   center    gas   in    this    line   \citep[hereafter
Paper~I]{paumard01}.   Data  were also  obtained  with  NIRSPEC  on Keck~II,  by
scanning the field  with the $24\arcsec$ slit used  in a north-south orientation
to obtain a  spectral cube covering 1.98~$\mu$m to  2.28~$\mu$m at resolution of
$\simeq 21.5$~km~s$^{-1}$ \citep{figer00}. With  the NICMOS cameras on board HST
the gas  was observed in another  infrared recombination line,  Pa$\alpha$, at a
spatial resolution of $0.18\arcsec$ \citep{scoville03}.  These data will be used
in this paper for comparison with the Br$\gamma$ data.

All  these data  show that  the ionized  gas  in the  inner few  parsecs of  the
Galactic  Center is  organized,  in projection,  into  a spiral-like  morphology
having several apparent  ``arms''.  This has led to  the widespread appellation,
``\object{Minispiral}'', for  this entire  pattern.  The brightest  features are
named ``\object{Northern Arm}'', ``\object{Eastern Arm}'', ``\object{Bar}'', and
``\object{Western  Arc}'', as  if it  imitates the  morphology of  a  very small
spiral  galaxy.   These  terms  seem  to  imply  that  the  ionized  filamentary
structures constituting Sgr~A West either  form spiral patterns, or are portions
of spiral arms.  This view was  motivated by the gas dynamical study carried out
by  \citet{lacy91},  who interpreted  the  [\ion{Ne}{ii}]  data  in terms  of  a
one-armed linear spiral in a Keplerian disk.  The various features of Sgr~A West
give a  spiral appearance primarily  because of the  way they are  superposed on
each  other.   However,  a  new   analysis  of  Lacy's  data  was  conducted  by
\citet{vollmer00}  to re-examine  the kinematic  structure of  the ionized  gas. 
Using a three-dimensional representation they confirm the standard features, but
with a  more complex structure,  including two features  for the Eastern  Arm: a
vertical finger  of high  density and a  large ribbon  extending to the  east of
\object{Sgr~A$^{\star}$}, and two distinctly different components in the Bar.

These  ISM  features   are  tidally  stretched  while  they   orbit  around  the
supermassive black  hole candidate  \object{Sgr~A$^{\star}$}, the exact  mass of
which, despite rapid progress, is still a matter of debate.  From stellar proper
motion      studies,      \citet{eisenhauer03}      give     a      mass      of
$3.6\pm0.6\times10^6$~M$_\odot$ and a  distance of $7.94\pm0.42$~kpc whereas the
estimate of \citet{ghez03a} is slightly higher at $4\pm0.3\times10^6$~M$_\odot$,
assuming   the   same   distance.    However,  \citet{aschenbach04}   derive   a
significantly lower  value from their study  of periodicity in  the black hole's
flares: $2.72^{+0.12}_{-0.19}\times10^6$~M$_\odot$.  On  the other hand, even if
the  supermassive black hole  is dominating  the potential  well in  the central
parsec,  the stellar  cusp \citep[e.g.][]{genzel03b}  may contain  a significant
fraction of the mass in the central arcsecond. \citet{mouawad2004} have explored
this possibility  and shown that the  stellar proper motion  data are consistent
with a  total mass  as high as  $4.8\times10^6$~M$_\odot$ for a  $25\%$ extended
component of the  mass distribution, the possible nature of  which they discuss. 
Therefore, fairly large error bars must still be put on the mass responsible for
the gravitational potential at the parsec scale.

In  the present  paper, the  gas content  in  the inner  region of  the GC  is
presented  and  analyzed from  high  spectral  resolution  data cubes  in  the
Br$\gamma$  and  \ion{He}{i}  2.06-$\mu$m  lines,  obtained  with  BEAR.   The
\ion{He}{i} data  are from  a new data  cube (larger field,  improved spectral
resolution)  compared  to  the   data  used  in  Paper~I.   A  multi-component
line-fitting procedure applied to the  emission-line profiles at each point of
the field  is described in  Sect.~\ref{sect:identproc}.  It was used  first on
the Br$\gamma$ cube and then on the \ion{He}{i} cube.  From this decomposition
in  Br$\gamma$, the identification  of defined  gas structures  comprising the
whole Sgr~A  West ionized region is presented  in Sect.~\ref{sect:structures}. 
Attempts  to adjust  Keplerian  orbits to  the  flowing gas  are presented  in
Sect.~\ref{sect:keplerian},  which  contains  in Sect.~\ref{sect:timescale}  a
discussion of the  implication of these identifications for  the formation and
the lifetime of the inner ionized gas.  All these elements allow us to discuss
the nature and origin of the ionized features in Sect.~\ref{discussion}. More
details on this analysis are given in \citet{paumardphdt}.

\section{Observations and preparatory data reduction}
The 3-D  data analyzed in  this paper were  obtained during two runs  with the
BEAR Imaging  FTS \citep{maillard95,maillard00} at the f/35  infrared focus of
the  3.6-m CFH  Telescope.  In  this  mode, a  $256\times256$ HgCdTe  facility
camera is  associated with the FTS,  in which several  narrow-band filters are
selectable.  Two  of them  were used, one  which contains the  Br$\gamma$ line
(4616.55~cm$^{-1}$,  bandpass  4585~--   4658~cm$^{-1}$)  and  the  other  one
centered  on  the  \ion{He}{i}  line at  4859.08~cm$^{-1}$  (bandpass  4806~--
4906~cm$^{-1}$).   The field of  view of  the instrument  is circular,  with a
diameter of  $24$\arcsec.  The Br$\gamma$ data  were acquired on  July 25, 26,
1997 (UT)  by observing  two overlapping fields  in order  to cover most  of a
field  of  $40\arcsec\times28\arcsec$, oriented  in  the East-West  direction,
centered  on  the position  of  Sgr~A$^{\star}$ (Fig.~\ref{tot_field}).   This
field contains  the Bar and  most of the  Northern and Eastern Arms,  but very
little of the Western  Arc.  The raw data consist of cubes  of 512 planes with
an integration time  of 7~s per image.  The maximum  path difference which was
reached in  the spectrometer determines the corresponding  limit of resolution
(FWHM) in velocity, equal in this study to $21.3$~km~s$^{-1}$.

On the following  night a single field centered  on Sgr~A$^{\star}$ was recorded
with  the $2.06$~$\mu$m  \ion{He}{i} filter.   The  analysis of  the later  high
resolution  data was  reported  in Paper~I,  which  brought new  results on  the
central  cluster  of  massive, hot  stars,  and  led  to  the detection  of  the
\object{Minispiral} in  helium.  However, the field  was not large  enough for a
significant areal coverage of the \object{Minispiral}.  New observations through
the same filter were therefore obtained on  June 9, 10, 11, 2000 in order to get
three overlapping circular  fields covering, when merged, most  of a total field
of $36\arcsec\times36\arcsec$, also  centered on Sgr~A$^{\star}$.  The estimated
width of the  interstellar $2.06$-$\mu$m line in Paper~I  called for an improved
spectral  resolution.    A  value  of  $\simeq   50$~km~s$^{-1}$  (exactly  FWHM
52.9~km~s$^{-1}$) was chosen instead of 74~km~s$^{-1}$ in the previous data, not
as high as  for Br$\gamma$, since the  line is weaker.  The raw  data consist of
cubes of 401 planes  with an integration of 20~s per image,  double the time for
the previous data, to improve the detection depth.

The processing  of the BEAR  data was described  in Paper~I; the main  steps are
standard  cube reduction,  atmospheric OH  correction and  correction  of filter
transmission  and   telluric  absorption  ---~particularly   important  for  the
2.06-$\mu$m data.   The various cubes  of the same  line are then merged  into a
mosaic.  The next step is the generation of the {\sl line cubes}, spectral cubes
in  which  the  continuum level  at  each  point  of  the  field is  fitted  and
subtracted, in  order to keep  only the emission  lines.  Figure~\ref{tot_field}
presents a three-color image obtained  from the Br$\gamma$ merged line cube, and
is a first glance at the velocity field of the region.

The Br$\gamma$  line cube is dominated  by the emission  from the interstellar
medium  (ISM),  but  some  stars  exhibit  the  Br$\gamma$  line  in  emission
(Fig.~\ref{tot_field}).   On the contrary,  in the  \ion{He}{i} line  cube the
stellar  emission from  the  hot  stars predominates  (Paper~I),  but the  ISM
emission is clearly detected too.

The central parsec was observed with the NICMOS cameras on board HST, during a
few runs between Aug.  1997 and  Aug.\ 1998, in 6 near-IR filters, including 2
narrow-band filters, F187N centered  on 1.87~$\mu$m Pa$\alpha$, and F190N, the
nearby  continuum.   By subtracting  the  F190N  filter  from the  other  one,
Pa$\alpha$  emission was  obtained  in a  field of  $19\arcsec\times19\arcsec$
centered  on Sgr~A$^{\star}$  (Stolovy 1999,  Scoville  {et al.}   2003) at  a
spatial resolution of $0.18\arcsec$, and  a wider field of $\simeq 120$\arcsec
at a  lower resolution of $\simeq  0.4\arcsec$.  We use an  image covering the
central $40\arcsec\times40\arcsec$  field from these  data for the  purpose of
comparison, in Fig.~\ref{PaumardT-fig5}.

%----------------------------------------------------------- 
\makeatletter\ifthenelse{\boolean{@referee}}{\begin{figure*}[]}{\begin{figure*}[!p]}
  \includegraphics[width=0.565\hsize]{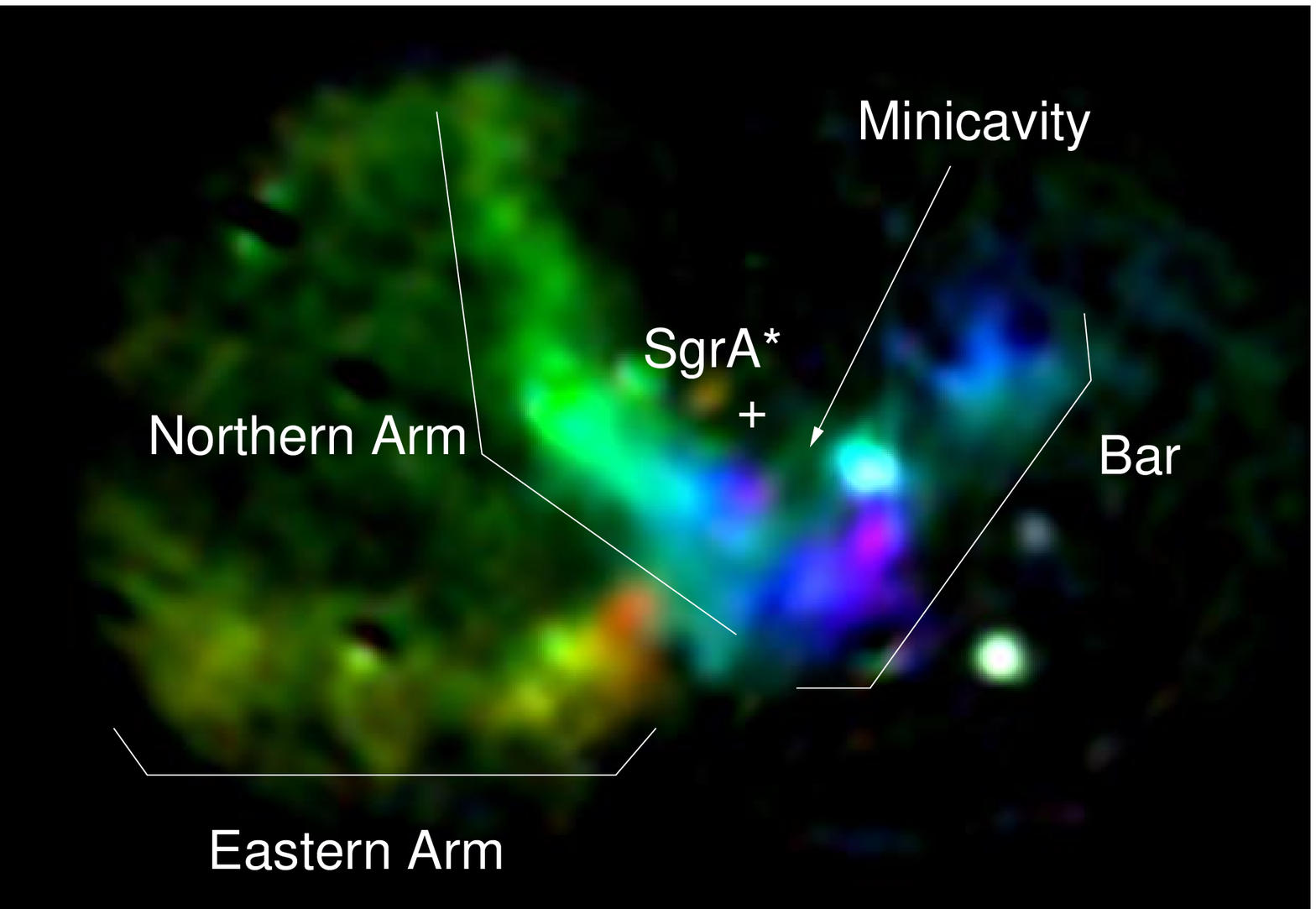}\hfill
  \begin{minipage}[b]{0.4\hsize}
    \caption{Three-color image of the two mosaicked    fields  of Sgr~A  West
      observed  with   BEAR  in   Br$\gamma$,  between  $-350$   (purple)  and
      $+350$~km~s$^{-1}$  (red).  The standard  bright features,  Northern and
      Eastern Arms, Bar, and  the \object{Minicavity}, are indicated.  Also, a
      few   emission  line   stars   show   up  as   bright   points  in   the
      image.\label{tot_field}}
    \end{minipage}
\end{figure*}
%-----------------------------------------------------------

\section{Structure identifications}
\label{sect:identproc}

At each  point of the field  the Br$\gamma$ and  \ion{He}{i} emission profiles
generally  appear complex.  The basic  assumption is  made that  each observed
profile results from the combination  of several velocity components, that is,
that along  any given line  of sight several  flows are superposed.  The first
goal of the  present paper is to separate these various  flows and to describe
them independently  from each other.  For  this purpose, the  development of a
multi-component line-fitting procedure able to  work on 3D data appeared to be
absolutely required.  From a coarse  examination of the datacube, fitting with
a maximum of four distinct velocity components along each line of sight seemed
adequate.

A comparison  of the velocity  components from one  line of sight to  the next
should usually reveal coherent velocity structures by continuity.  In the end,
it might  be possible  to conclude whether  these structures are  isolated, or
form continuous flows.   The process is thus split into  two main parts: first
the line profile decomposition at all the points of the field, and second, the
structure identification.   This work is based on  original software developed
by  Miville-Desch\^enes  (personal   communication),  which  we  have  largely
extended.

\subsection{Line profile decomposition}

\subsubsection{Line profile}

A  single velocity  component of  the  emission lines  from the  ISM has  been
assumed  to be  a  Gaussian $I(v)$,  with  three free  parameters: $I_0$,  the
amplitude         of         the         Gaussian         expressed         in
$\mathrm{erg}\cdot\mathrm{s}^{-1}\cdot\mathrm{cm}\cdot\mathrm{pixel}^{-1}$
($1\ \mathrm{pixel}=0.125\ \mathrm{arcsec}^2$);  $v_0$, the radial velocity of
the component; and $\Sigma$, the width  of the line, due to thermal agitation,
turbulence  and any  velocity gradient  along the  line of  sight or  across a
resolution element.   The full  width at  half maximum (FWHM)  of the  line is
given by: $ \mathrm{FWHM}=2\Sigma\sqrt{2\ln2} $.

The instrumental line shape of the FTS is a sinc function:
\begin{equation}
\psi(v)            =           
\frac{\sin\left(\pi\,{\delta_m}\,\frac{v\sigma_0}{c}\right)}
                   {\pi\,{\delta_m}\,\frac{v\sigma_0}{c}}
\end{equation}
where $\sigma_0$  is the  central wavenumber and  $\delta_m$ the  maximum path
difference between  the two arms  of the interferometer, which  determines the
limit of resolution d$\sigma$ of the data with d$\sigma=0.6/\delta_m$ (FWHM).

The  measured line  profile  is  thus the  convolution  product $S=I*\psi$,  a
function  of three  free parameters  $I_0$, $\Sigma$  and $v_0$.   Each single
spectrum  of the  field has  been fitted  to a  set of  four such  lines, thus
implying twelve free parameters.

\subsubsection{Procedure}
\label{procedure}
\paragraph{Preparation:}  As for any  fitting routine,  a reasonable  initial
guess must be provided for each point of the field.  For such a problem, where
we  intend to fit  complex line  profiles at  low signal-to-noise,  the method
cannot be fully automatic.  The operator freely chooses a few starting points,
for which he  is able to provide an  unambiguous decomposition. These starting
points should be  chosen so that every structure in  the field is represented,
and keeping  in mind that the most  complex regions are better  fitted if they
are close to  a starting point. In  our case, five starting points  were used. 
From the starting  points, a first procedure attempts  to fit a four-component
line  shape function to  each spectrum.   For each  new spectrum,  the initial
guess is  determined from the results  found for the  neighboring points.  The
spectra are studied sequentially in parallel spiral-mode scannings around each
starting point.   Except for the initial  guess, the fitting of  a spectrum is
independent of all the others.
\paragraph{Step~1:} The velocity structures are then built. A totally
determinist program finds  the brightest point in the  field, and examines its
neighbors, searching for  a component such that the  velocity gradient between
the point of interest and the neighbor is less than a certain amount, which is
set by the operator at runtime. If several velocity components of the examined
neighbor spectrum satisfy this velocity gradient criterion, the component with
the  highest  amplitude  is  assigned  to the  structure.   The  procedure  is
iterative, and once a few neighbors have been selected into a structure, their
neighbors  are  in turn  examined  for  possible  selection.  A  structure  is
considered to  be complete when  it cannot be  extended any further,  that is,
when no  component of  the spectra  lying at its  border satisfy  the gradient
criterion.  Construction  of the next structure is  then undertaken, beginning
with the brightest point  in the data cube not yet selected  into a structure. 
The procedure  stops when every component  of every spectrum in  the field has
been assigned  to exactly one  spatial structure.  This procedure  allows only
one velocity component at any location  to be selected into a given structure,
and conversely  each component at any  location can be selected  into only one
structure.   A structure  that  overlaps itself  spatially,  thus causing  two
velocity  components  along  the  same  line  of  sight,  cannot  be  directly
identified as such: the program splits it into two structures.
\paragraph{Step~2:} The structures containing less  than a  given number of 
points, chosen by  the operator, are discarded.  To  be validated, a structure
must be more extended than a spatial resolution element, which yields a region
more  extended  than  a $3\times3$  pixel  box  in  our case.   After  several
attempts, we  reached the conclusion  that no structure containing  fewer than
$50$  pixels was significant  in our  field: if  such structures  exist, their
overall signal-to-noise  ratio is less  than $3\sigma$.  The  discarded fitted
lines  corresponding to these  abortive structures  are either  stellar lines,
noise spikes, or actual ISM lines under the detection threshold.
\paragraph{Step~3:}  This procedure requires that the  detected structures be
visually inspected.   The operator then has  the possibility to  add some more
common sense heuristics into  the structure identification, a little difficult
to implement but easy to apply manually. Several problems can occur:
  \begin{itemize}
  \item  the line fitting  procedure might  fit only  one component  where two
    blended components are indeed more appropriate;
  \item  during  step~2,  if  two  overlapping  structures  intersect  in  the
    $\alpha$--$\delta$--velocity    space,   the    procedure    can   falsely
    cross-connect them, i.e.  reconstruct  two structures, each one being made
    of  parts  of  both  physical structures  (Fig.~\ref{crossconnect}).   The
    second  derivative of  the  velocity map,  as  well as  continuity of  the
    intensity map, can  be used by the operator to  decide which connection is
    the good one.  Preliminary work  has been done to automatically take these
    criteria into account.
    \begin{figure}[!t]
      \includegraphics[width=\hsize]{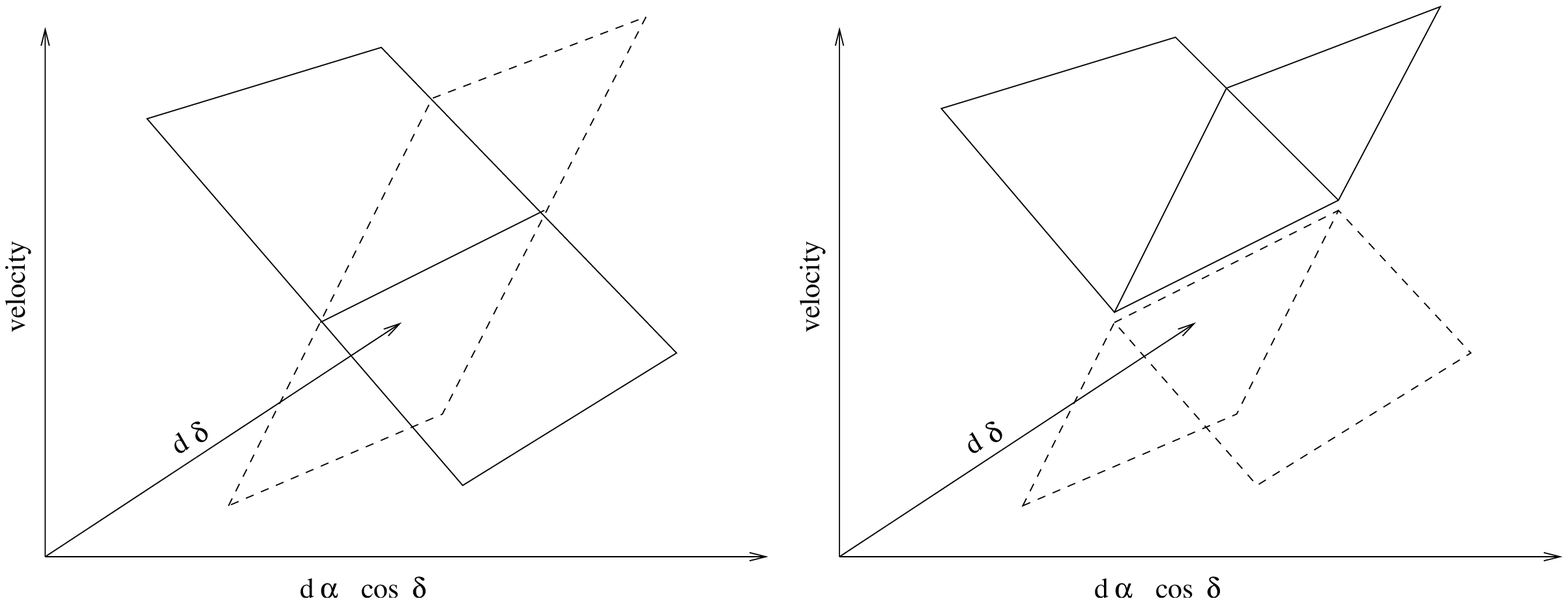}
      \caption{\label{crossconnect}Cross-connection  problem.    Left:  two  physical
        structures  (solid and dashed  polygons) intersect  each other  in the
        $\alpha$--$\delta$--velocity    space.     Right:    the    structures
        reconstructed by the software can be erroneous.}
    \end{figure}
  \end{itemize}
  It is also possible at this  time to interpolate the results if they contain
  holes, and  to extrapolate them over a  few pixels in order  to provide good
  initial guesses for the next step.
\paragraph{Step~4:}  Next, these  manually  corrected  results  are used  to
perform a second fit at each point of the field; at this point, 2D information
is entirely included  in the initial guess provided to  the fitting procedure. 
Since components  have been discarded  during step~2, not all  initial guesses
still contain four components.  Steps~1 to 4 can be iterated a number of times
to reach a stable result, in our case eight times.

\section{Results}
\label{sect:structures}

\subsection{General description of the results}
\begin{figure*}
\makeatletter\ifthenelse{\boolean{@referee}}{}{\hrule}
\vspace{0.5\baselineskip}
\begin{center}\includegraphics[width=0.9\hsize]{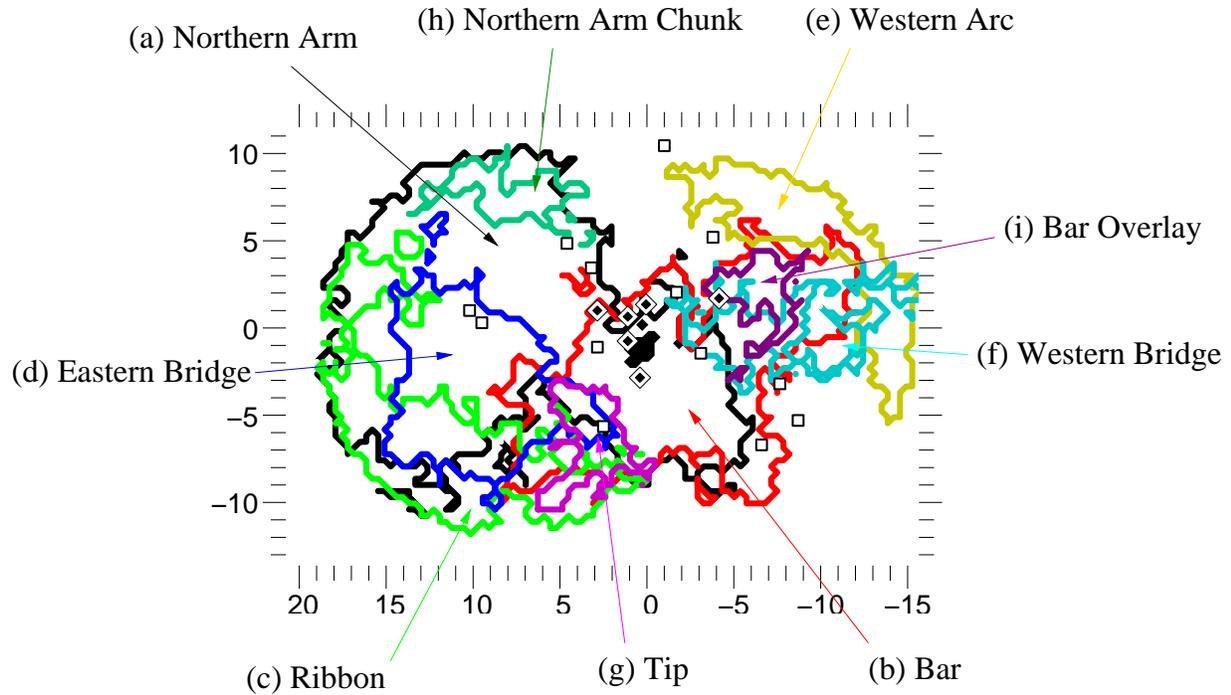}\end{center}
\begin{minipage}{0.40\hsize}\begin{center}\includegraphics[bb=74 386 602 905,width=\hsize]{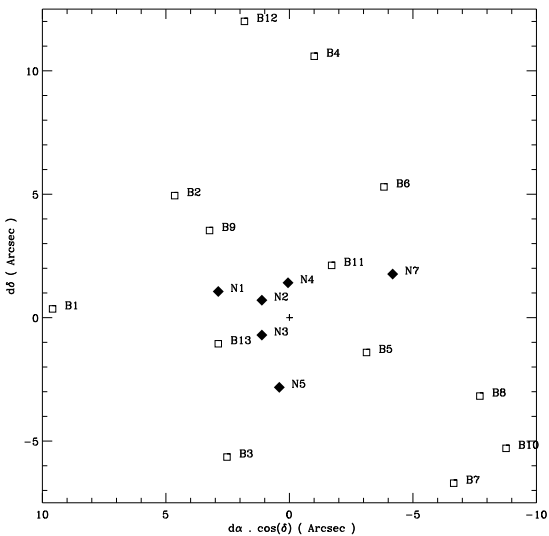}\end{center}\end{minipage}\hfill
\begin{minipage}{0.55\hsize}\caption{\label{ism-stars}Above:  outline  of 
    every structure.  The region filled in black corresponds to the points where
    two lines associated with the \object{Northern Arm} are detected (see text).
    Filled diamonds (resp.  empty  squares) represent narrow (resp.  broad) line
    helium stars.   Left: IDs of  the narrow (N)  and broad (B) line  stars, see
    table below for  common-names identification \citep[Paper~I,][and references
    therein]{paumard03}.\vspace{\baselineskip}}
\makeatletter\ifthenelse{\boolean{@referee}}{\resizebox{\hsize}{!}{%
\begin{tabular}{ll|ll|ll}
\hline\hline
ID & Name & ID & Name & ID & Name \\
\hline
N1 & \object{GCIRS 16NE}  & B1 & \object{ID~180}   & B7 & \object{AF star}          \\
N2 & \object{GCIRS 16C}   & B2 & \object{GCIRS 7E2}  & B8 & \object{AF NW}        \\
N3 & \object{GCIRS 16SW}  & B3 & \object{GCIRS 9W}   & B9 & \object{HeIN3}       \\
N4 & \object{GCIRS 16NW}  & B4 & \object{GCIRS 15SW} & B10 & \object{BSD WC9}    \\
N5 & \object{GCIRS 33SE}  & B5 & \object{GCIRS 13E2} & B11 & \object{GCIRS 29N}    \\
N7 & \object{GCIRS 34W}   & B6 & \object{GCIRS 7W}   & B12 & \object{GCIRS 15NE}   \\
                            &&&& B13 & \object{GCIRS 16SE2}  \\
\hline
\end{tabular}
}}{%
\begin{tabular}{ll|ll|ll}
\hline\hline
ID & Name & ID & Name & ID & Name \\
\hline
N1 & \object{GCIRS 16NE}  & B1 & \object{ID~180}   & B7 & \object{AF star}          \\
N2 & \object{GCIRS 16C}   & B2 & \object{GCIRS 7E2}  & B8 & \object{AF NW}        \\
N3 & \object{GCIRS 16SW}  & B3 & \object{GCIRS 9W}   & B9 & \object{HeIN3}       \\
N4 & \object{GCIRS 16NW}  & B4 & \object{GCIRS 15SW} & B10 & \object{BSD WC9}    \\
N5 & \object{GCIRS 33SE}  & B5 & \object{GCIRS 13E2} & B11 & \object{GCIRS 29N}    \\
N7 & \object{GCIRS 34W}   & B6 & \object{GCIRS 7W}   & B12 & \object{GCIRS 15NE}   \\
                            &&&& B13 & \object{GCIRS 16SE2}  \\
\hline
\end{tabular}
}\makeatother
\end{minipage}
\end{figure*}
Both the Br$\gamma$ and \ion{He}{i}  data have been analyzed with the software
described above.  This  leads to a vision of the  \object{Minispiral} more complex than
usually  thought, one which  is consistent  with, but  more detailed  than the
description proposed by \citet{vollmer00}.  After a careful examination of the
Br$\gamma$ data we identify 9 components  of various sizes, labeled (a) to (i)
(Fig.~\ref{ism-stars}).  A  description of these structures,  and their radial
velocity and  flux maps, are  presented in Appendix~\ref{maps}.  Two  types of
velocity  maps appear,  some  with a  significant  overall velocity  gradient,
others without any appreciable,  large-scale velocity gradient.  The deviation
from the  mean motion,  defined as the  local difference between  the velocity
measured  at one  point and  the mean  value for  the neighboring  points, and
divided  by the  uncertainty ($3\sigma$  error bars  from  the multi-component
line-fitting  procedure), ranges from  roughly one  tenth to  ten for  all the
features, which  means that every  velocity structure shows  significant local
features.

The areal size of the  structures (Table~\ref{tablesurf}), expressed in terms of
solid angle covered on the  sky, ranges from 17~arcsec$^2$ to 300~arcsec$^2$ for
the \object{Northern Arm}. The surface area of each structure must be considered
as a lower limit because BEAR may  not detect the weakest parts, and because the
field of view does not cover the entire \object{Minispiral}.
\begin{table*}[htb]
  \caption{Feature identifications, with surface areas (pixels
    and  square   arcseconds),  and   minimum  and  maximum   radial  velocities
    (km~s$^{-1}$).  The  last column  gives  [\ion{He}{i}]/[Br$\gamma$] for  the
    different      structures,     relative      to      the     mean      value
    $<$[\ion{He}{i}]/[Br$\gamma$]$>$.  The  Minicavity  is  separated  from  the
    \object{Northern Arm}, as it warrants special attention. \label{tablesurf}\label{table:lr}}
  \begin{minipage}{\textwidth}
    \begin{center}
      \begin{tabular}{clrrrrc}
\hline\hline
ID & Feature name & S (pix) & S (arcsec$^2$) & Vmin & Vmax & [\ion{He}{i}]/[Br$\gamma$]\footnote{normalized to its mean value}\\
\hline
a&\object{Northern Arm} & 2414 & 300.8 & -290 & 186 & 0.74\footnote{except
  \object{Minicavity}} \\
& \object{Minicavity} &&&&& 0.85 \\
b&\object{Bar} & 1389 & 173.1 & -214 & 194 & 0.99 \\
c&\object{Ribbon} & 833 & 103.8 & 130 & 240 & 0.78 \\
d&\object{Eastern Bridge} & 670 & 83.5 & 32 & 180 & 1.09 \\
e&\object{Western Arc} & 471 & 58.7 & -40 & 72 & 0.52 \\
f&\object{Western Bridge} & 327 & 40.7 & -124 & 98 & 1.73 \\
g&\object{Tip} & 207 & 25.8 & 220 & 336 & 2.64 \\
h&\object{Northern Arm Chunk} & 185 & 23.1 & 12 & 72 & -- \\
i&\object{Bar Overlay} & 136 & 16.9 & -270 & -10 & 1.81 \\
\hline
      \end{tabular}
    \end{center}
  \end{minipage}
\end{table*}

\label{sect:heibrg}
The same  work of  decomposition into velocity  structures on  the \ion{He}{i}
data  was more  difficult  than for  the  Br$\gamma$ data  since the  spectral
resolution and signal-to-noise ratio are lower.  The fact that the \ion{He}{i}
data are dominated  by the emission from the helium  stars also contributes to
the greater complexity  of this task. We thus skipped  the preparatory step of
the decomposition  process, and  provided directly a  complete set  of initial
guesses based on the Br$\gamma$ results, since at first sight the distribution
of ionized gas is globally the same in \ion{He}{i}.  This method precludes the
\ion{He}{i} analysis from being fully  independent, although steps~1 to 4 were
performed eight times, until the procedure converged satisfactorily.

Finally, all the  structures detected in Br$\gamma$ are  detected in \ion{He}{i}
as well, except the \object{Northern Arm  Chunk} (h), which is probably too weak
in  this  line.   However,  several  differences  in  the  appearance  of  these
structures  are  noticeable,  and  detailed  in  Appendix~\ref{app:heibrg}.   To
quantify these differences, we  have built [\ion{He}{i}]/[Br$\gamma$] line ratio
maps for  each structure, normalized to the  areal mean for this  ratio over the
union of  all the structures.  The [\ion{He}{i}]/[Br$\gamma$]  line ratio varies
considerably across the field, so  that, for instance, the \object{Northern Arm}
bright rim  and the  Minicavity do not  show the  same shape in  \ion{He}{i} and
Br$\gamma$ (Fig.~\ref{hei_brg}).  The  last column of Table~\ref{table:lr} shows
the  mean normalized  [\ion{He}{i}]/[Br$\gamma$]  line ratio  for the  different
structures.  It  appears that this  ratio is lower  than the mean value  for the
main, well-known features --~the \object{Northern Arm}, the \object{Ribbon}, the
\object{Bar}  and  the  \object{Western  Arc}~--  and  higher  for  the  smaller
features.  However, the  values are computed only for  the positions detected in
both Br$\gamma$ and \ion{He}{i}, so they  do not take into account the faintest,
least excited regions of each feature.
\begin{figure}
  \includegraphics[width=\linewidth]{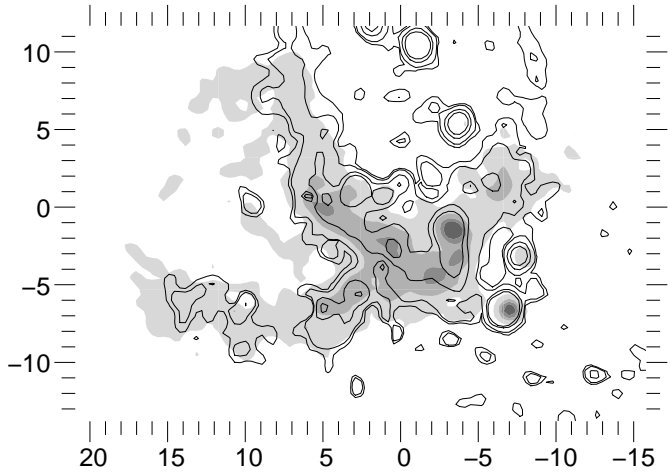}
\caption{Comparison between integrated flux in Br$\gamma$ (grey scale) and
  \ion{He}{i} (empty  contours).  Axes are  offsets from Sgr~A$^\star$
  in arcsec.\label{hei_brg}}
\end{figure}

\subsection{Brief description of the main features}

Although all  nine structures are thoroughly  described in Appendices~\ref{maps}
and  \ref{app:heibrg}, we  briefly summarize  here the  most important  results. 
Contrary to its standard description, the \object{Northern Arm} is not seen here
as a bright N-S lane, but as  an extended, triangular surface.  One edge of this
triangle is the bright rim generally noticed, but it extends all the way over to
the Eastern Arm.   The third edge of the  triangle is the edge of  the field, so
viewing  this  feature  in  a  larger  field  may  yield  a  somewhat  different
description.  It  contains the Minicavity.   The [\ion{He}{i}]/[Br$\gamma$] line
ratio is higher on the western side of the bright rim of the Northern Arm and on
the inner side of the Minicavity than in the rest of the structure.

As already  described by  \citet{vollmer00}, the Eastern  Arm region  is split
into two  parts: a  \emph{Ribbon} (c) and  a \emph{Tip}  (g). For the  sake of
clarity,  we chose  not to  name  any feature  ``Eastern Arm''.  At the  elbow
between  the  Ribbon  and  the  Tip,  in  the  \object{GCIRS~9W}  region,  is  a
bubble-like  feature, or  a \emph{\object{Microcavity}}  (radius $\simeq1\arcsec$),
with  a rather  bright rim  (Fig.~\ref{bulle}),  which appears  at a  specific
velocity (230~km~s$^{-1}$).  Finally, a  secondary rim parallel to the Eastern
Arm (Fig.~\ref{PaumardT-fig5}, from spot~A eastwards) is often associated with
it.  Here, we  see that this feature belongs to  an independent structure, the
Eastern Bridge (d).
\begin{figure}
  \resizebox{\hsize}{!}{\includegraphics{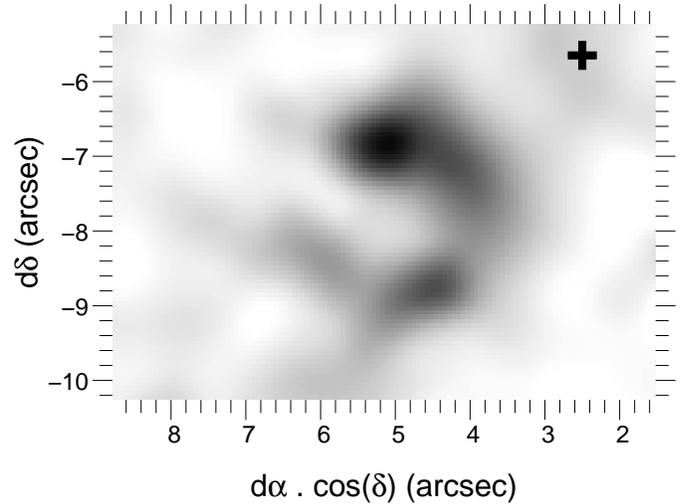}}
  \caption{\object{Microcavity}  feature   in  the  region   of  GCIRS~9W, represented
    by  the black  cross.  Axes  are offsets  from  Sgr~A$^\star$.  Integrated
    velocity range: $220$--$240$~km~s$^{-1}$.
    \label{bulle}}
\end{figure}

\subsection{Extinction by the structures}
\label{extinction}

One  of the most  difficult questions  concerning the  ionized features  is to
determine their  relative positions, i.e.,  when two  structures overlap, which
one is closer to the observer.  On two occasions, the flux maps can be used to
infer this information.
  
The flux map  of the {Northern Arm} (Fig.~\ref{nafmap}) shows  a region of low
intensity, the northwestern boundary of which is a well defined line, oriented
approximately  northeast--southwest.   This  intensity discontinuity  is  most
obvious south of \object{GCIRS~1} and west of the Minicavity.  This line follows
very closely the outline of  the Eastern Bridge. It goes from $\simeq2\arcsec$
east  and  $5\arcsec$  south  to  $13\arcsec$ east  and  $3\arcsec$  north  of
Sgr~A$^\star$, where it becomes more  difficult to locate precisely because of
a lower signal-to-noise ratio.  If we interpret the intensity discontinuity in
terms  of  an  increment of  the  extinction,  this  gives  us two  pieces  of
information:  the   Northern  Arm  is   behind  the  Eastern  Bridge   on  the
line-of-sight, and the  Eastern Bridge contains a substantial  amount of dust,
responsible for the  extinction of about $50\%$ of the  Br$\gamma$ flux of the
Northern  Arm,  or  about  $\simeq0.76$  magnitudes  at K.   The  ratio  of  K
extinction  to  visual  extinction  being  about $0.1$,  the  inferred  visual
extinction is around $7.6$ magnitudes.   Using the area of the Eastern Bridge,
83.5  arcsec$^2$   (Table~\ref{tablesurf}),  and  the   conversion  factor  of
$2\times10^{21}$~cm$^{-2}$ per visual magnitude of extinction, we estimate the
mass of the Eastern Bridge  to be $\simeq15$~M$_{\odot}$.  This estimate is to
be considered  with caution  since the conversion  factor may be  different in
this extreme environment, and since we have only one measurement point for the
extinction, although the Br$\gamma$ emission  seems to correlate well with the
extinction  map  inferred  from  the  Northen Arm's  emission  map,  which  is
difficult to interpret. Furthermore, this estimate does not include a possible
extension of  the Eastern Bridge  outside our field  of view.  However,  it is
comparable to  the mass  of $\simeq27$~M$_\odot$ estimated  by \citet{liszt03}
for the Bar.

Although the flux map of the Bar (Fig.~\ref{barfmap}) is less smooth than that
of the {Northern Arm}, making such effects more difficult to see, the shape of
the periphery  of the Minicavity is  clearly identified in  extinction on this
map,  showing that the  Bar is  behind the  {Northern Arm}  along the  line of
sight.

\section{Keplerian orbit fitting}

\label{sect:keplerian}

The velocity maps give a view of the features very different from the usual flux
maps which,  by themselves, can be  misleading.  For instance  the morphology of
the \object{Northern Arm} with its typical rim may lead one to think of this rim
as the  true path for most  of the material,  whereas the velocity map  does not
show any peculiar  feature at the location of  the rim.  We are thus  led to the
idea that the  kinematics of the {Northern Arm}  should be studied independently
of its intensity  distribution. To do so, we have tried  to analyze the Northern
Arm as  a Keplerian system,  the location and  mass of the central  object being
those  of Sgr~A$^{\star}$.   Its position  with respect  to  \object{GCIRS~7} is
taken  from  \citet{menten97}, and  the  positions of  the  stars  in the  field
relative to \object{GCIRS~7}  from \citet{ott99}.  The distance to  SgrA* is the
value  of   8~kpc  reported  by  \citet{reid93}.    We  have  used   a  mass  of
$3\,10^6$~M$_\odot$ for  most of our models,  and we will discuss  the impact of
changing this mass later.

\subsection{Fitting one orbit on a velocity map}

For a  first, simple approach, we  created a dedicated IDL  graphical package. 
With this  tool, the  operator can  easily adjust one  Keplerian orbit  over a
velocity map. A  Keplerian orbit in 3D is defined  by five orbital parameters:
the eccentricity,  two angles defining  the orientation of the  orbital plane,
the periapse  (distance of closest  approach to the  center of motion),  and a
third angle defining the position of the periapse.

Once the  operator is  almost satisfied with  the orbital parameters  found by
trial and error, an automatic fitting  procedure can be called. It is possible
to  fix parameters,  and the  orbit can  be forced  to go  through  a selected
constraint  point  by  tying  the  periapse to  the  other  parameters.   Good
agreement can be  found between observed and calculated  velocities, except in
the region of the Minicavity, so we attempted to model the full velocity field
of  the Northern  Arm with  a bundle  of Keplerian  orbits bounded  by various
constraints.   We note  that  this model  alone  is not  sufficient to  decide
whether the  orbits are bound  to the gravitational  field of the  black hole,
since the  bound and unbound  solutions are not dramatically  different within
our field of view.

\subsection{Fitting a bundle of orbits on a velocity map}
\label{fitting-bundle}

To fit several orbits  at a time on a velocity map, we force each one to
pass  through  a  different constraint  point,  as  in  the  one-orbit  case,  
with the constraint points chosen to be on  different physical orbits.  The 
result described  here uses 50 constraint points, evenly 
spaced on  the solid  line of Fig.~\ref{namap}.   Another constraint  line has
been tried  as well (dashed  line), with consistent results.   Each constraint
point is  given an  index, increasing from  the point  nearest Sgr~A$^{\star}$
outwards, that is used to refer to a given orbit.

To ensure a smooth model --~we are interested only in the global motion~-- the
four functions that map each constraint point to one of the orbital parameters
have  been chosen to  be described  as spline  functions, uniquely  defined by
their value at a number of control points, chosen among the constraint points.
The number of  points used to define the spline function  can be freely chosen
to set  the spatial resolution  of the model  across the flow.   After several
attempts,  we have  chosen  to fix  this number  to  four in  our final  model
(yielding a resolution of $8.8\arcsec$).  Thus, having four functions (one for
each of  the orbital parameters), each  of them being defined  by four values,
the model depends on sixteen parameters.

We designed  a fitting procedure  to adjust this  model based on  the observed
velocity  map  by  minimizing  the  reduced  \mbox{$<\!\chi^2\!\!>$}.   It  is
possible to  either fix  some parameters, or  to force  them to have  the same
value for each orbit.  This way,  for example, it is possible to check whether
the observed velocity  map is consistent with coplanar  orbits or with uniform
eccentricity. To avoid  studying only local minima in  the parameter space, it
is also important to use several initial guesses.

\subsection{Homothetic hypothesis}
We again designed  an IDL graphical tool to easily study  whether the data are
consistent with a homothetic\footnote{Two orbits are said to be ``homothetic''
  when they  are identical except for  their scale, i.e.  when  they share the
  same  orbital parameters,  except the  periapse distance.}   set  of orbits,
which is the simplest model.

With  the hypothesis of  a homothetic  set of  orbits, the  eccentricity still
cannot  be well  constrained.   Bound orbits  seem  to be  preferred, but  the
agreement is as good with circular  orbits and very eccentric orbits, close to
parabolic.   The  residual  map  always  has  the  same  shape:  the  observed
velocities are always  smaller than the computed ones along  the inner edge of
the bundle of  orbits, and higher along the outer  edge.  The global agreement
is always poor, with \mbox{$<\!\chi^2\!\!>^{1/2}\simeq 70$}.

\subsection{General case}
A few of these homothetic models have been chosen as initial guesses for other
adjustments, with released constraints.  It  is first interesting to check the
coplanar hypothesis, in which only  the two parameters that define the orbital
plane  are  kept  uniform,  and  the  uniform  eccentricity  hypothesis.   The
agreement is much  better when leaving either the  eccentricity or the orbital
plane free. In the following, both parameters are free.

Even in the  most general situation, the parameters  are still not constrained
enough  to extrapolate  the model  outside  the field  of view,  or to  derive
reliably  the direction of  proper motion.   However, the  models share  a few
characteristics that we judge to be robust because of their repeatability: the
orbital planes are close to that of the \object{CND}; the orbits are not quite
coplanar, the two angles that define  the orbital plane varying over a $\simeq
10^\circ$  range; the  eccentricity  varies  from one  orbit  to another,  the
innermost orbits  being hyperbolic ($e$ of  the order of 2)  and the outermost
closer to circular ($e\lesssim 0.5$).

\begin{figure*}
\includegraphics[width=0.32\textwidth,bb=21 13 321 236]{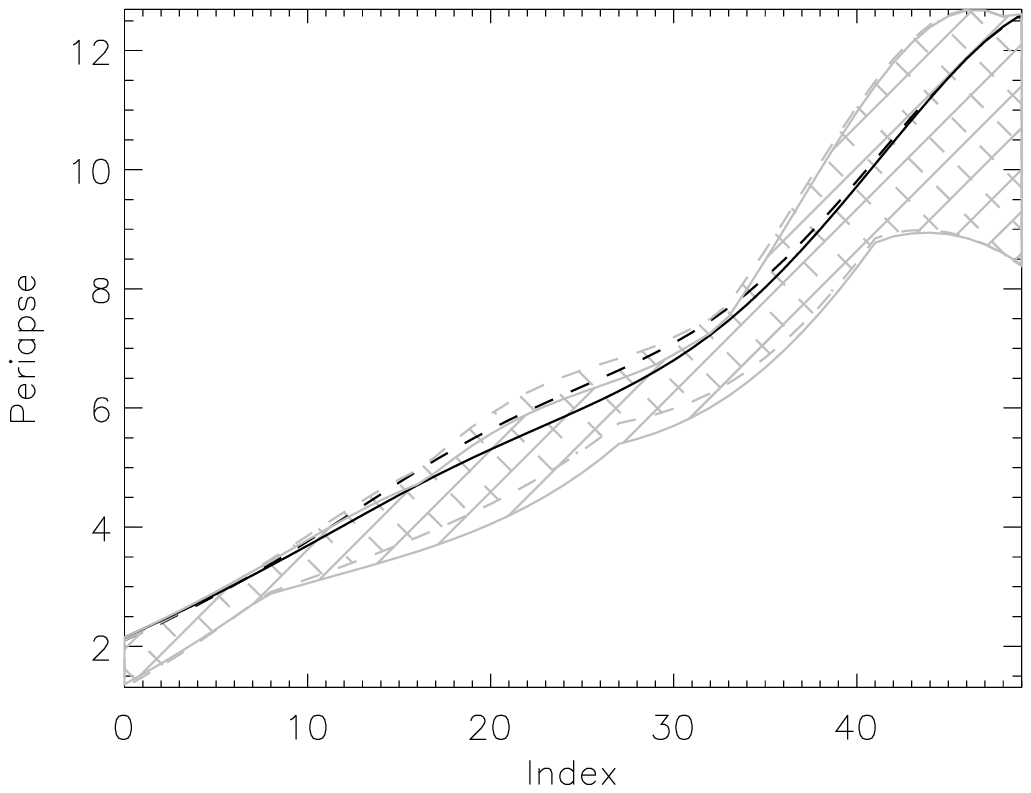}
\includegraphics[width=0.32\textwidth,bb=21 13 321 236]{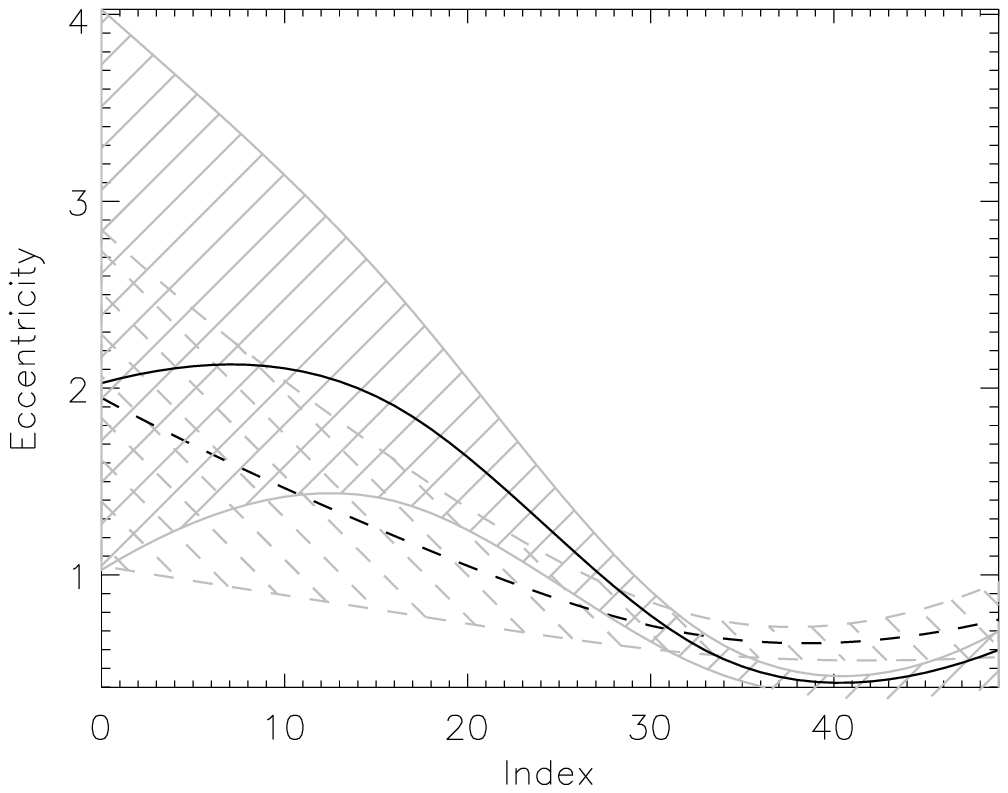}
\includegraphics[width=0.32\textwidth,bb=21 13 321 236]{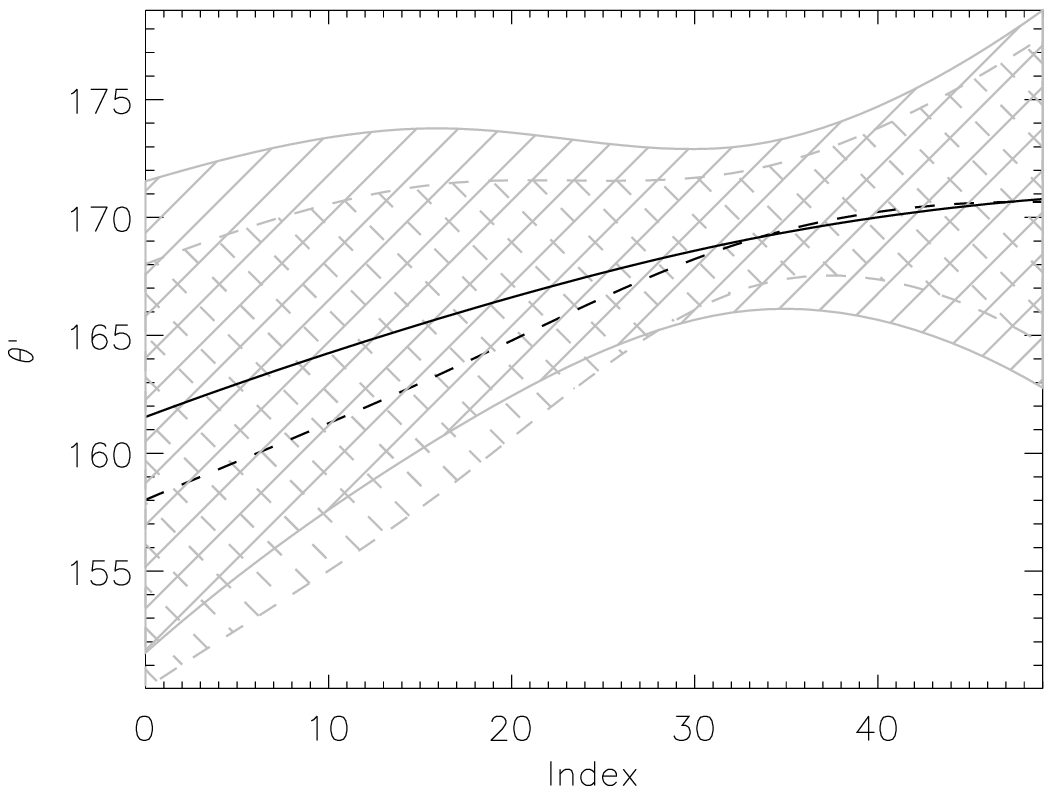}\linebreak
\begin{minipage}{0.32\textwidth}\includegraphics[width=\textwidth,bb=21 13 321 236]{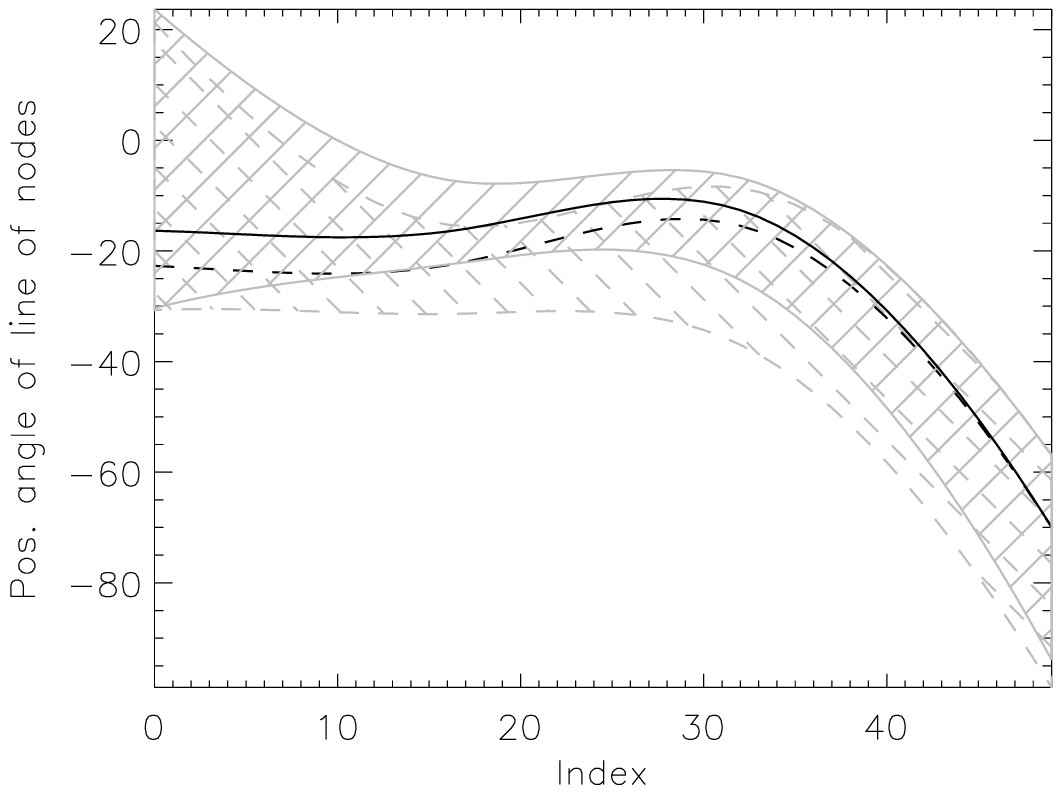}\end{minipage}
\begin{minipage}{0.32\textwidth}\includegraphics[width=\textwidth,bb=21 13 321 236]{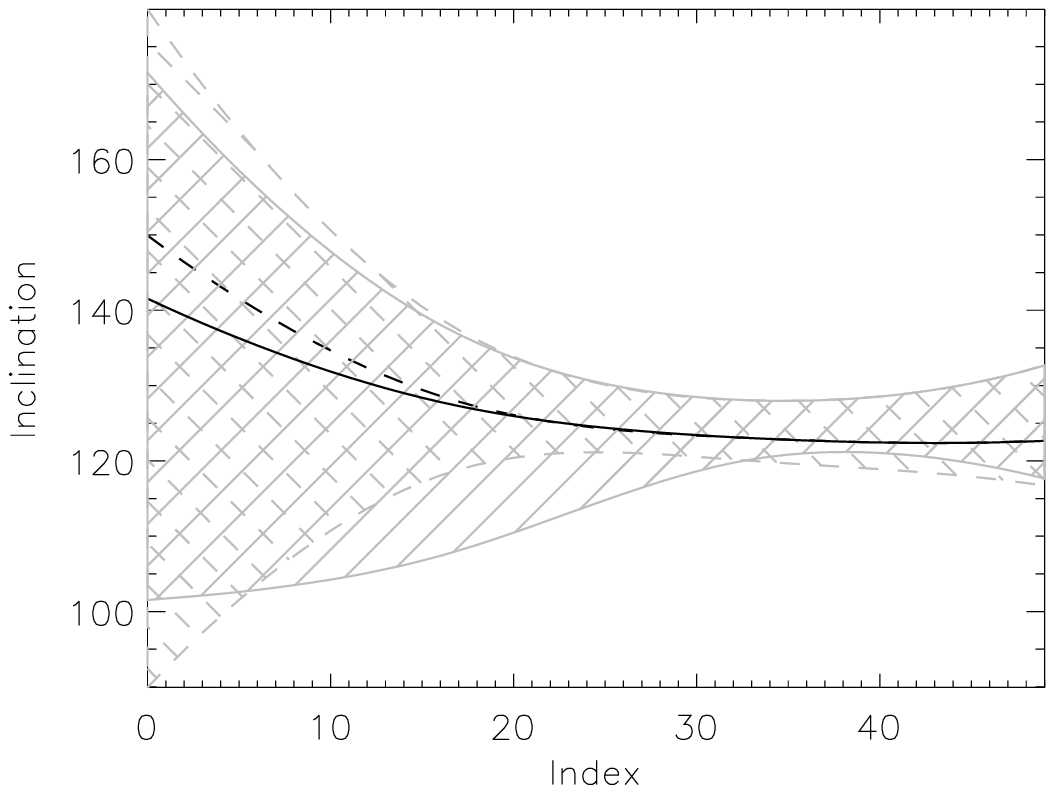}\end{minipage}
\begin{minipage}{0.32\textwidth}
\caption{\label{bestmodel:parameters}These  five plots  show  the laws
  of each of the five orbital  parameters in our two best models. Solid lines:
  $M_\mathrm{SgrA*}=3\,10^6\,\mathrm{M}_\odot$.          Dashed         lines:
  $M_\mathrm{SgrA*}=4\,10^6\,\mathrm{M}_\odot$.    The  gray   hashed  regions
  correspond to the $2\sigma$ error bars.  Angles are in degrees, the periapse
  in equivalent arcsec on the sky.}
\end{minipage}\linebreak\null

\end{figure*}

\subsection{3D morphology and time-scale of the \object{Northern Arm}}
\label{sect:timescale}

\begin{figure*}
  \resizebox{\textwidth}{!}{\includegraphics{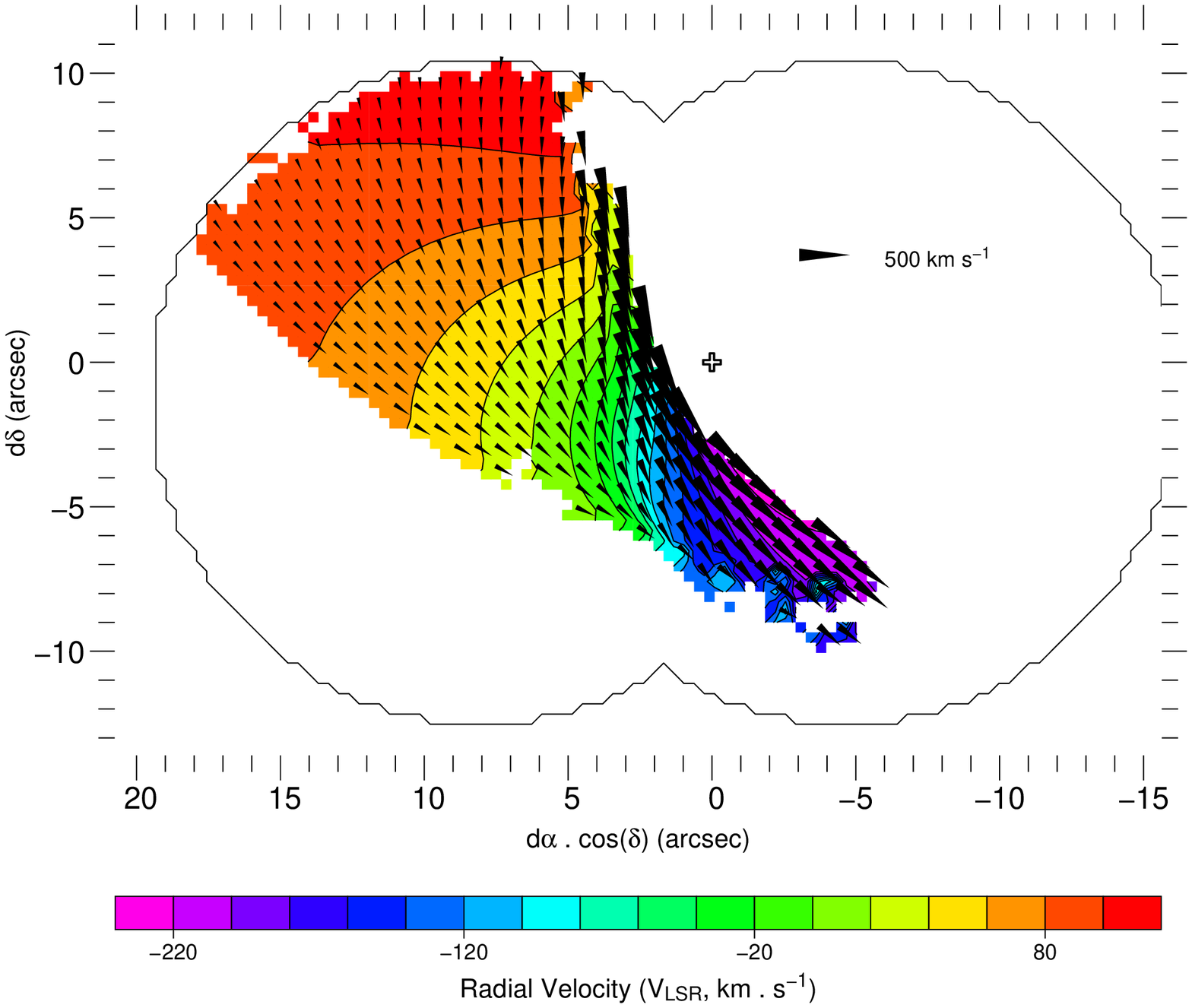}}
  \resizebox{0.56\textwidth}{!}{\includegraphics{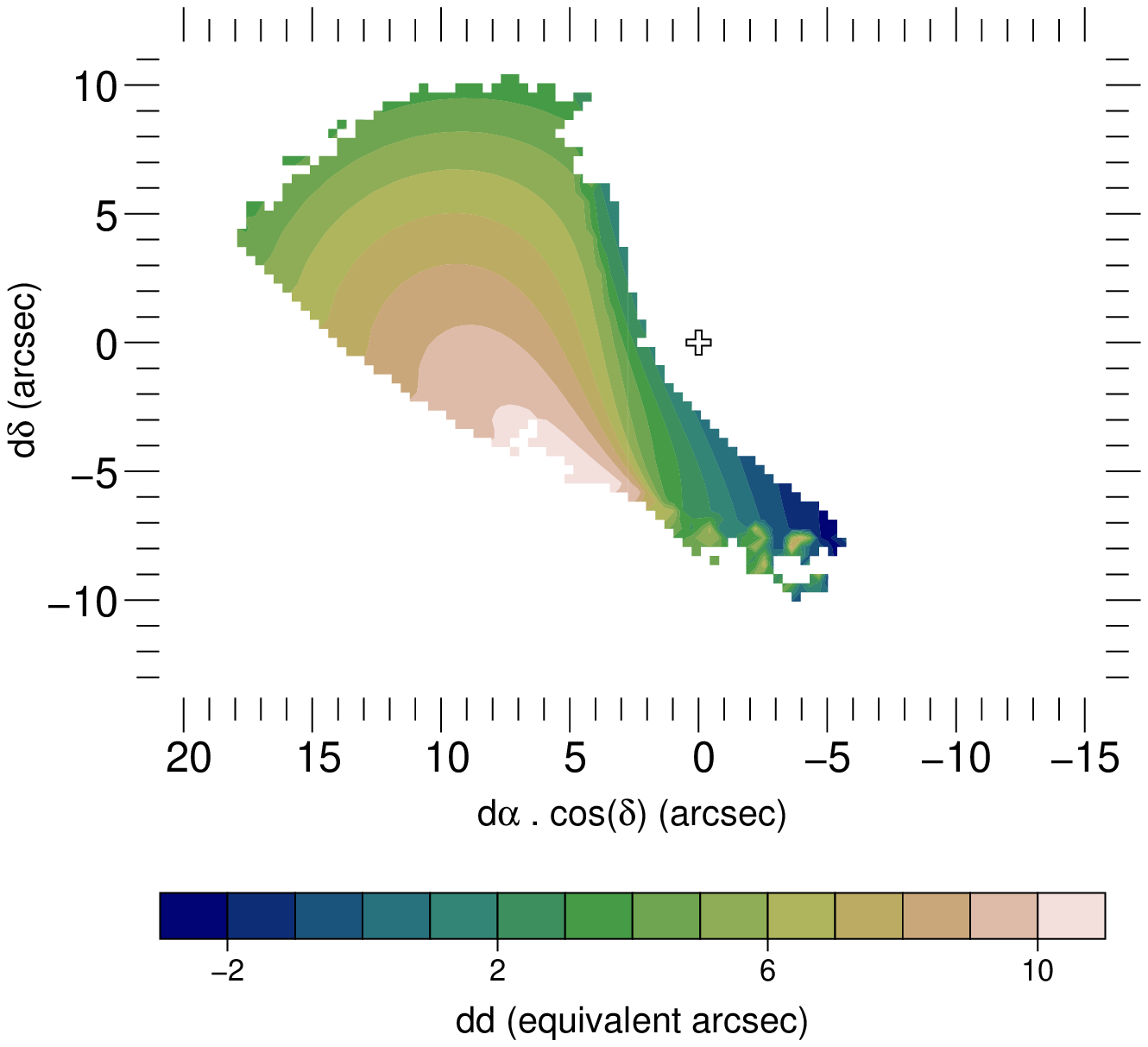}}\hfill
  \begin{minipage}[b]{0.4\textwidth}
    \caption{Above: 3D velocity map for the \object{Northern Arm} from our
      model.      The      arrows     show     the      derived     tangential
      velocities.\label{vmap3d}\label{elevation}\label{fits-vmap}         Left:
      elevation  map for  the detected  part of  the {Northern  Arm}  from our
      model, given in equivalent  arcseconds for homogeneity.  d$d$ stands for
      differential of the distance, positive  d$d$ means further away from the
      observer  than  the  center  of  mass,  Sgr~A$^\star$,  at  d$\alpha=0$,
      d$\delta=0$, d$d=0$.}
  \end{minipage}
\end{figure*}
We   present  here   our  best   model,  i.e.,   the  one   with   the  lowest
\mbox{$<\!\chi^2\!\!>^{1/2}$} among  the realistic  models that cover  most of
the   {Northern  Arm}.   The   laws  used   for  this   model  are   shown  in
Fig.~\ref{bestmodel:parameters}. The agreement between the radial velocity map
of    this    model    and    the    observed   velocity    map    is    good:
\mbox{$<\!\chi^2\!\!>^{1/2}=18$}.   The  histogram   of  the  radial  velocity
difference  between  our  model  and  measurements  is  close  to  a  Gaussian
distribution centered  on zero with $\sigma=10$~km~s$^{-1}$.   That means that
the  method  is  unbiased,  and   that  the  mean  error  is  10~km~s$^{-1}$.  
Figure~\ref{elevation},  also available at  CDS\footnote{via anonymous  ftp to
  cdsarc.u-strasbg.fr             (130.79.128.5)             or            via
  http://cdsweb.u-strasbg.fr/cgi-bin/qcat?J/A+A/} in FITS format, shows the 3D
velocity map  of the \object{Northern Arm}  derived from this  model. The FITS
version\label{online-data} consists  of four FITS  files. The first  gives the
geometry of the cloud  by means of the distance to the  observer at each point
of the field; the  second contains a stack of three maps,  each one giving one
velocity component.  The two other files give the error bars for the two first
files.

The variations  of the orbital parameters  induce a particular 3D  shape for the
{Northern Arm} (Fig.~\ref{PaumardT-fig5}): for  all the non-coplanar models, the
{Northern Arm} looks like a saddle-shaped surface, and this warped shape induces
a crowding of orbits that closely  follows the bright rim of the structure.  Two
different geometries could explain this  saddle shape: the Northern Arm could be
a warped planar structure, or have  approximately the shape of the inner side of
a torus.  An infalling neutral cloud, tidally stretched by the black hole, would
have such a torus-like geometry.  If, in addition, its inner side was ionized by
hot  stars  located   still  further  inside,  near  the   black  hole,  as  the
\object{GCIRS~16} stars are, one must expect this ionized skin to have precisely
the same saddle  shape as the Northern  Arm.  The bright rim itself  is not only
due to the  stronger UV field and  a real local enhancement of  the density, but
also to an enhancement of the column density due to the warping.  An interesting
point is  that, in some  models, \emph{no} orbit  follows the bright  rim, which
emphasizes that  it is really  important to consider the  dynamics independently
from the  morphology of the  {Northern Arm}.  Another characteristic  present in
all models is that the period of  the orbits ranges from a few $10^4$~years to a
few $10^5$~years, which implies that  the {Northern Arm} would have a completely
different  shape in  a few  $10^4$ years,  and cannot  be much  older  than that
time-scale.  These  results are  in good agreement  with the conclusions  of the
model  of  \citet{sanders1998} for  explaining  the  observed  structure of  the
ionized  gas filaments, in  particular the  Northern Arm,  explained in  term of
disruption of gas clouds in the potential of a point mass.

Since the agreement  in radial velocity is now rather good,  it makes sense to
look at  the deviations from global motion  by looking at the  features on the
residual velocity map (Fig.~\ref{PaumardT-fig5}):
\begin{description}
\item[A)] the  flow shows  a rather significant  deviation in the  region just
  southwest of the embedded  star, \object{GCIRS~1W}; this perturbation could be
  due to the interaction with the wind  of this star; indeed a bow shock where
  the wind meets the ambient Northern  Arm gas has been invoked to explain the
  observed morphology of \object{GCIRS~1} \citep{tanner04}.
\item[B)]  the region  of  this model closest  to the Minicavity  is
  perturbed;
\item[C)] another  deviation is seen at  the precise location  where the bright
  rim bends abruptly, just east of \object{GCIRS~7E2};
\item[D)] finally, an elongated feature is seen on the fainter rim coming from
  \object{GCIRS~1W} towards the northeast.
\end{description}

\begin{figure*}
\includegraphics[width=0.48\textwidth,bb=21 46 680 675]{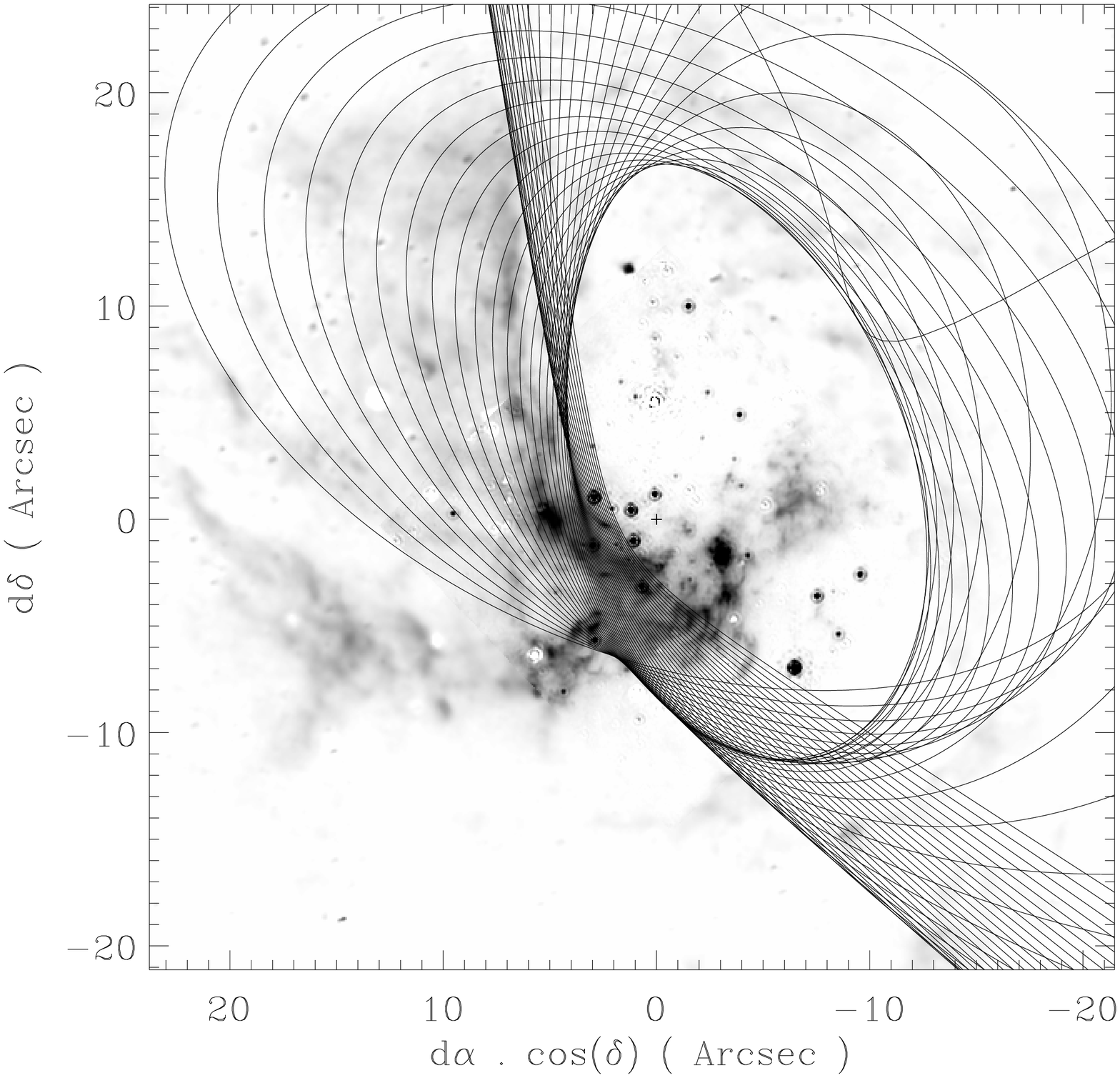}\textbf{a)}\hfill
\includegraphics[width=0.48\textwidth,bb=21 46 680 675]{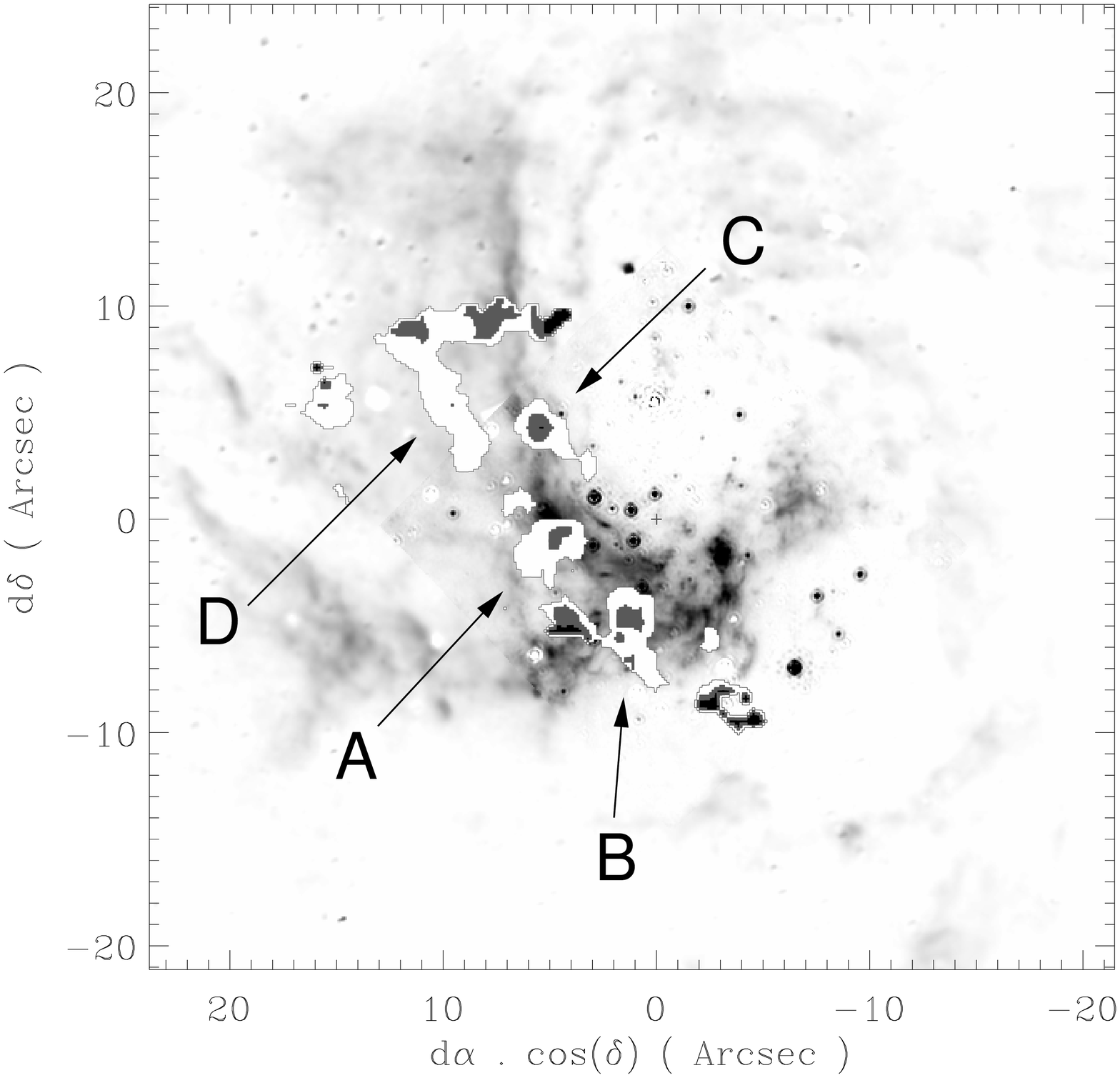}\textbf{b)}
\caption{\label{PaumardT-fig5}\textbf{a)} on this Pa$\alpha$ map \citep{scoville03}, one
  of the  Keplerian models is  overplotted.  \textbf{b)} the  most significant
  deviations from Keplerian  motion discussed in the text are  labeled A to D,
  and indicated as filled contour.}
\end{figure*}

\section{Discussion}
\label{discussion}

The geometry  of the  \object{Northern Arm} has  been studied from  its velocity
map,  leading to  the conclusion  that it  may not  be planar,  but rather  is a
three-dimensional  structure.  The  best  model discussed  above,  which uses  a
central mass of $3\times10^6$~M$_\odot$, has  been used as the initial guess for
another  adjustment  using  $4\times10^6$~M$_\odot$.   With this  value  of  the
central    mass,   every   remark    made   above    is   still    valid.    The
\mbox{$<\!\chi^2\!\!>$}  is slightly  degraded, but  not significantly  changed. 
This analysis cannot provide stringent constraints on the central mass.

Figure~\ref{PaumardT-fig5}a  is quite  compatible with  the Northern  Arm indeed
being   the   ionized  surface   of   a   neutral   cloud  \citep[as   suggested
by][]{jackson93,telesco96}.  This figure also suggests that the Northern Arm and
the Western Arc  may be two parts of the same  physical structure.  The velocity
derived from  the model  agrees with  the measured velocity  of the  Western Arc
within $\simeq50$~km~s$^{-1}$ (which is reasonable since it is an extrapolation)
and has the right gradient, but this  coincidence is lost when a central mass of
$4\times10^6$~M$_\odot$ is used.   However, the adjustment is not  made over the
entire Northern Arm,  since our field of  view is limited.  The same  study on a
complete map  of the Northern Arm  would probably show whether  the Northern Arm
and Western Arc are one and the same physical feature.  In any case, the orbital
plane of the Northern Arm is very  close to that of the CND, which suggests that
the Northern Arm may  have originated from a cloud in a  stable orbit inside the
CND, which would have been  extracted from there through a cloud-cloud collision
for instance.   However, from their study  of proper motions of  features in the
\object{Minispiral}, \citet{yusef98} show that some material of the Northern Arm
seems to be  on hyperbolic orbits, which we confirm, and  which implies that the
captured  clouds  lose  some  mass  in  their process  of  evolution  into  long
filaments.    The    tangential   velocity    field   of   the    Northern   Arm
(Fig.~\ref{vmap3d}) is  interestingly similar to  the magnetic field  derived by
\citet{aitken98}, which is easily explained by the shear in our model.

We have  shown that at least  two structures are thick,  dusty clouds, because
their extinction factor  is of the order of several  $10\%$ at $2$~$\mu$m: the
Eastern Bridge and  the edges of the Minicavity in the  Northern Arm. From the
wider  field in  Pa$\alpha$,  we can  assume  that the  Eastern  Bridge is  an
east-west elongated cloud,  of which only the western part is  in the field of
view of BEAR. If this cloud is moving westwards along its principal axis, this
western part  must be  the leading side  of the  cloud along its  orbit, which
would explain the lack of shear inferred from its velocity map.

The presence of  three isolated ionized gas structures  (the Western Bridge, the
Northern Arm Chunk and the Bar  Overlay) in addition to the standard large flows
and  to  the  Eastern  Bridge,  which   seems  to  be  another  flow,  has  been
demonstrated.  Some of  these structures may be isolated gas  patches, but it is
also  possible that some  of them  are regions  of the  neutral clouds  in which
ionized fronts form the \object{Minispiral}, locally excited.  For instance, the
Bar Overlay, the velocity map for which  is very similar to that of the Bar, may
be a  region belonging  to the  same neutral cloud  as the  Bar, which  would be
locally  excited  by  \object{GCIRS~34W}.  The  [\ion{He}{i}]/[Br$\gamma$]  line
ratio is significantly  higher for these tenuous features  than for the standard
Northern  Arm,  Eastern  Arm  and  Bar.   This ratio  is  variable  across  each
structure, which  can basically be explained in  two ways: first, it  can be the
trace of  local enrichment of the  gas in helium, and  second, it can  be due to
local enhancements of  the excitation, either because of a  stronger UV field or
of shocks.

There  are  about  $20$ high  mass  loss  stars  in  the region  (Paper~I  and
references therein).   A typical  mass loss rate  for stars of  these spectral
types    is    of    the    order    of    $\simeq10^{-4}$~M$_\odot$~yr$^{-1}$
\citep{najarro94}.   This material  must reside  in the  central parsec  for a
duration similar to the time-scale of the Northern Arm: $\simeq10^4$~yr.  From
these considerations, the total mass of interstellar gas in the central parsec
coming from the  mass loss of these stars  must be around a few  tens of solar
masses.  On the  other hand, if the ionized structures  are really the ionized
fronts of  neutral clouds, these clouds could  have a mass similar  to that of
the    clouds    that    form    the    CND:    $\simeq10^3$~M$_\odot$    each
\citep{christopher03}.   It would then  be unlikely  that interstellar  gas of
stellar origin  contribute significantly  to the enrichment  of these  clouds. 
However, such  a significant contribution remains plausible  if the individual
structures are indeed much less massive than $10^3$~M$_\odot$, as suggested in
Sect.~\ref{extinction}.

There is  a clear correlation  between the projected  proximity of gas  to the
helium stars and the [\ion{He}{i}]/[Br$\gamma$] line ratio:
\begin{itemize}
\item  two  of  the  gas  patches   detected  in  both  lines  having  a  high
  [\ion{He}{i}]/[Br$\gamma$]  ratio  are   coincident  with  the  helium  star
  \object{GCIRS~34W};
\item the Bar, which  is the main feature with the highest  line ratio, is close
  to  the  \object{GCIRS~16} helium  star  cluster,  and  seems to  contain  the
  \object{GCIRS~13E} star cluster  (Appendix~\ref{morphology}), which is made of
  several high-mass-loss-rate stars \citep{maillard03};
\item the  Tip, the feature  with the highest  [\ion{He}{i}]/[Br$\gamma$] ratio,
  seems  to   be  interacting  with  a   stellar  wind,  as   evidenced  by  the
  \object{Microcavity}, and  is on  the same  line of sight  as the  helium star
  \object{GCIRS~9W};
\item  the differences  in  the shape  of  the Northern  Arm  between the  two
  spectral lines appear  to come from the geometry of the  UV field around the
  \object{GCIRS~16} cluster.
\end{itemize}
The discrepancies that are most difficult  to explain are the high brightness in
\ion{He}{i},  in contrast  to  their  relative faintness  in  Br$\gamma$ in  the
southwestern parts  of both the Minicavity  and the Tip.  However,  this part of
the Minicavity is  rather close in projection to the  \object{AF star}, the flux
of which, if  this proximity is not only in  projection, could favor \ion{He}{i}
emission.  Finally, our results  remain consistent with interstellar material of
uniform composition, distributed  in a non-uniform UV field,  the exact value at
any given point depending on the 3D localization of nearby hot stars.

In addition to  this, a \object{Microcavity} has been  discovered at the elbow
between  the Eastern Arm  Ribbon and  Tip.  It  is probably  a new  example of
interaction between stellar wind or polar jet and an ISM cloud, similar to the
Minicavity.    Another  apparent  star-ISM   interaction  phenomenon   is  the
interaction of  the Northern  Arm gas flow  with \object{GCIRS~1W} (spot  A in
Fig.~\ref{PaumardT-fig5}b).  These interactions show  that the dynamics of the
flows must  be influenced by the  stars, as \citet{yusef93}  suggested for the
wind of the \object{GCIRS~16} cluster.

\section{Conclusion}

We begin to  gain access to the relative positions of  features along the line
of sight: the Eastern Bridge is  closer to the observer than the Northern Arm,
and the Bar is behind the Northern  Arm. In addition to that, knowledge of the
radial  velocity  field of  the  Northern  Arm has  allowed  us  to propose  a
kinematic  model, which  provides a  three dimensional  map of  this  feature. 
Having such maps for all of the  ISM features would give us the opportunity to
estimate  the UV  field that  hits these  ISM features,  taking  the shadowing
effects into  account.  It would then  become possible to  estimate the helium
abundance in the different structures from their relative line ratios.  and to
assess the  notion that these  structures have incorporated  enriched material
processed through the post-main sequence  stages of evolution of massive stars
and then ejected as winds.

This work has  been performed on a  field covering most of the  inner parts of
the  \object{Minispiral}.   However, repeating  the  same  analysis  on a  wider  field
containing  the \object{Minispiral} to  its full  extent, would  allow one  to directly
check whether  the Northern Arm and  the Western Arc are  related features, an
important  constraint on  their formation  scenario.  Moreover,  obtaining the
velocity  maps of  a  wider field  would  allow one  to  better constrain  the
parameters  of the  Keplerian fit  to the  Northern Arm,  and may  then reveal
deviations   from  the   Keplerian  model,   due  to   momentum  loss   or  to
non-gravitational  forces.  This  would  be  a very  interesting  clue to  the
accretion  process.  This  program requires  a wide-field  spectro-imager with
spectral and spatial resolutions comparable to those of BEAR.

\begin{acknowledgements}
  We  are  grateful   to  M.A.~Miville-Desch\^enes\footnote{Currently  at  the
    Canadian Institute for Theoretical Astrophysics (Toronto -- Canada).} from
  the \emph{Institut d'astrophysique spatiale} (Orsay -- France) for giving us
  the original version  of the IDL spectral decomposition  package and helping
  us  in  the early  stages  of customizing  and  expanding  it.  Mark  Morris
  acknowledges support from  the CNRS for a stay at the  IAP during which this
  collaboration progressed.   Also, Mark Morris's  participation was partially
  supported by NSF grant AST-9988397.
\end{acknowledgements}

\bibliographystyle{aa} \bibliography{0209}

% The appendices will be online-only. Need last version of aa.cls (as of writing): 5.3
% Nope, they're finally going to be printed
% \Online

\appendix
\section{Morphology of the ionized gas in Sgr A West}
\label{maps}
\label{fmaps}
\label{morphology}
A description  based on  the analysis  of the Br$\gamma$  data follows  for each
identified  velocity  structure.  Their  Br$\gamma$  line  flux (peak  intensity
$\times$  width   of  the   line)  and  radial   velocity  maps  are   given  in
Figs.~\ref{namap}  to  \ref{bovmap}.    Axes  show  offsets  from  Sgr~A$^\star$
(represented as a cross), in arcseconds.   The line flux maps are the gray scale
images, left-hand column; a black outline  gives the full extent of the detected
structure. The velocity  maps are the color level  images, right-hand column; to
improve  contrast,  the  color  scale  is  not  the  same  for  each  map.   The
binocular-shaped black line shows approximately the field boundaries.

(a) \emph{\textsf{\object{Northern Arm} (Fig.~\ref{namap}):}}  It extends from its well
known bright N-S lane all the way  over to the Eastern Arm.  The third edge of
this triangularly shaped feature is the  edge of the field. Its flux is partly
absorbed (Sect.~\ref{extinction}) by the  Eastern Bridge, the outline of which
is shown  as a dashed line  on the Northern Arm  flux map.  As  it reaches the
Minicavity,  the {Northern  Arm}  is split  into  two layers  in the  spectral
direction: on the spectra of a few adjacent pixels (14), two lines are clearly
detected, indicating two layers of gas that both connect continuously with the
rest  of the  Northern Arm  (each layer  is separately  found by  our software
described  Sect.~\ref{procedure}).   The  main   layer  contains  all  of  the
Minicavity,  while the second  layer seems  to be  deflected northward  of the
Minicavity,  and  forms  the  small   finger  between  the  two  helium  stars
\object{GCIRS~16SW} (N3)  and \object{GCIRS~33SE} (N5,  Fig.~\ref{ism-stars}).  It
extends  further away  by $\sim5\arcsec$  to the  northwest, and  contains the
point-like  feature  just  above   the  aperture  of  the  Minicavity  (source
$\epsilon$ of \citealt{yusef90}).   On the few pixels where  both features are
detected, the secondary layer  is $50$--$80$~km~s$^{-1}$ more blueshifted than
the   main  one.    Both  layers   are  represented   in  the   velocity  maps
(Figs.~\ref{vmap3d} and \ref{namap}), indicating the velocity of the secondary
one for the few points outlined in red, where both are detected.  The flux map
(Fig.~\ref{nafmap}) gives  the sum  of the two  layers.  The direction  of the
Northern  Arm   motion  (from  north   to  south)  has  been   established  by
\citet{yusef98}.   The  kinematics  of   the  {Northern  Arm}  are  thoroughly
discussed  in  Sect.~\ref{sect:keplerian}.   In  the  velocity  map,  the  two
straight lines  represent the constraint  lines used for our  Keplerian models
(Sect.~\ref{fitting-bundle}).

(b)  \emph{\textsf{\object{Bar} (Fig.~\ref{barmap}):}}  It is  very extended  (from the
Ribbon of  the Eastern  Arm (c) to  the Western  Arc (e)), very  straight, and
shows a smooth overall velocity gradient.   Its flux is partly absorbed by the
edges of the Minicavity (Sect.~\ref{extinction}),  a few contours of which are
given as dotted  lines in the flux map  of the Bar.  The map  peaks sharply at
the location of the  \object{GCIRS~13E} compact star cluster \citep{maillard03},
which seems  to show that this object  excites local material in  the Bar, and
thus  must  be  either embedded  in  it,  or  very  close  to it.   Thus,  the
coincidence between this bright spot and  the northern end of the western edge
of  the  Minicavity  seems to  be  a  projection  effect,  and not  physical.  
\citet{vollmer00} mention two complementary  components of the Bar, which they
call  \object{Bar~1}  and  \object{Bar~2},  though their  description  is  not
sufficient  to  determine  precisely  the  positions of  these  two  suggested
components.  We also see two additional features, which we propose to call the
{\sl Western Bridge} (f) and {\sl Bar Overlay} (i).  Parts of the Bar are also
superimposed  on almost  every other  structure, including  the Ribbon  of the
Eastern Arm  (c), the Tip (g), the  Eastern Bridge (d) and  the {Northern Arm}
(a).

(c) \emph{\textsf{\object{Ribbon} (Fig.~\ref{eamap}):}}  As already described by
\citet{vollmer00},  the  Eastern   Arm  region  is  split  into   two  parts:  a
\emph{Ribbon}  and a \emph{Tip}  (g).  The  velocity gradient  of the  Ribbon is
directed along  the minor  axis of the  structure, not  along its major  axis as
expected for a flow.

(d) \emph{\textsf{\object{Eastern Bridge} (Fig.~\ref{ebmap}):}}  A structure of
medium size  extends from the  Ribbon (c) to  the bright rim of  the {Northern
  Arm}.  It  does not  show any large-scale  velocity gradient, and  its shape
does  not  show  any  principal  axis  that would  indicate  a  flow.   It  is
superimposed  on  the  faint  regions   of  the  {Northern  Arm},  and  partly
superimposed  on the  Ribbon,  the Bar  and  the Tip.   Its  southern side  is
parallel to, as well as superimposed  upon, the Ribbon; the two structures are
probably  related, although  their  relative velocities  differ  by more  than
50~km~s$^{-1}$.  The name we propose is based on the fact that it lies between
the two Arms  of the \object{Minispiral}, both in the  spatial and spectral dimensions,
being close to  the Ribbon in the spectral dimension on  its southern side and
to the {Northern Arm}  on its northern side.  It is also  inspired by the fact
that  the most  luminous part  of it  in our  field is  a small  vertical bar,
seemingly connecting the  bright parts of the Northern  and Eastern Arms. This
bar is located around $7\arcsec$ east  of Sgr~A$^\star$, and extends from about $1$
to   $6\arcsec$   south   of    Sgr~A$^\star$.    However,   the   Pa$\alpha$   map
(Fig.~\ref{PaumardT-fig5})  shows  that  this   bar  may  extend  outside  our
field-of-view into  an elongated  feature parallel to  the Ribbon,  going from
about  $7\arcsec$ east  and $1\arcsec$  south to  $\simeq23\arcsec$ east  and $9\arcsec$  north of
Sgr~A$^\star$.  The lack  of an overall gradient in  the velocity map suggests
that this  feature is  not much affected  by shear.   It seems related  to the
entity formed by the  combination of the Ribbon and the Tip;  it may belong to
it, or be interacting with it.

\textsf{(e)  \emph{\object{Western Arc}  (Fig.~\ref{bomap}):}}  The Western  Arc
lies just  at the edge  of the field,  so we have  access only to  its innermost
part.   It  is seen  as  a  rather simple  feature,  with  large scale  velocity
gradient.  It is  superimposed  on the  Western  Bridge on  a  few pixels.   The
velocity  field that  we measure  is basically  in good  agreement with  that of
\citet{lacy91}.

\textsf{(f)  \emph{\object{Western  Bridge}  (Fig.~\ref{wbmap}):}}  The  Western
Bridge is a tenuous, elongated feature oriented east-west and extending from the
Bar to the Western Arc.  This structure,  as well as the Bar and the Bar Overlay
(i)  upon which  it  is superimposed,  contains  in projection  the helium  star
\object{GCIRS~34W} (N7, Fig.~\ref{ism-stars}).

\textsf{(g) \emph{\object{Tip} (Fig.~\ref{tipmap}):}} The Tip is, in projection,
a very concentrated  and relatively small object with  the most redward velocity
in the  region ($\simeq300$~km~s$^{-1}$).  The  Tip has already been  noticed by
\citet{vollmer00} only on  a morphological basis, as a  finger-shaped portion of
the Eastern Arm  in their three-dimensional data.  Here, we  see that the Ribbon
and the  Tip are two  distinct features, superimposed  on the line of  sight (we
detect  two  lines on  $72$  lines-of-sight),  and  separated by  a  \object{Microcavity}
(Fig.~\ref{bulle}),   thus  we   do  not   adopt   the  representation-dependent
denomination ``\object{Finger}''.

\textsf{(h)  \emph{\object{Northern Arm  Chunk}  (Fig.~\ref{nacmap}):}} A  small
tenuous structure is  seen superimposed on the {Northern  Arm}, a few arcseconds
north of \object{GCIRS~7}.  It lies at the edge of our field, so it could extend
further out; however  the Pa$\alpha$ image shows a small,  horizontal bar at its
location, crossing the  bright rim of the {Northern Arm}, that  does not seem to
be much extended.

\textsf{(i)  \emph{\object{Bar Overlay}  (Fig.~\ref{bovmap}):}} The  Bar Overlay
looks like a small cloud that is superimposed upon the western region of the Bar
and that  shows a velocity gradient  similar to the one  of the main  Bar at the
same  location, with an  offset of  $\simeq-40$~km~s$^{-1}$.  This  may indicate
that these  two features are closely  related.  They could, for  example, be the
two faces of a single neutral cloud, ionized by two distinct UV sources.

\begin{figure*}
  \begin{minipage}[b]{0.47\hsize}
    \resizebox{\hsize}{!}{\includegraphics{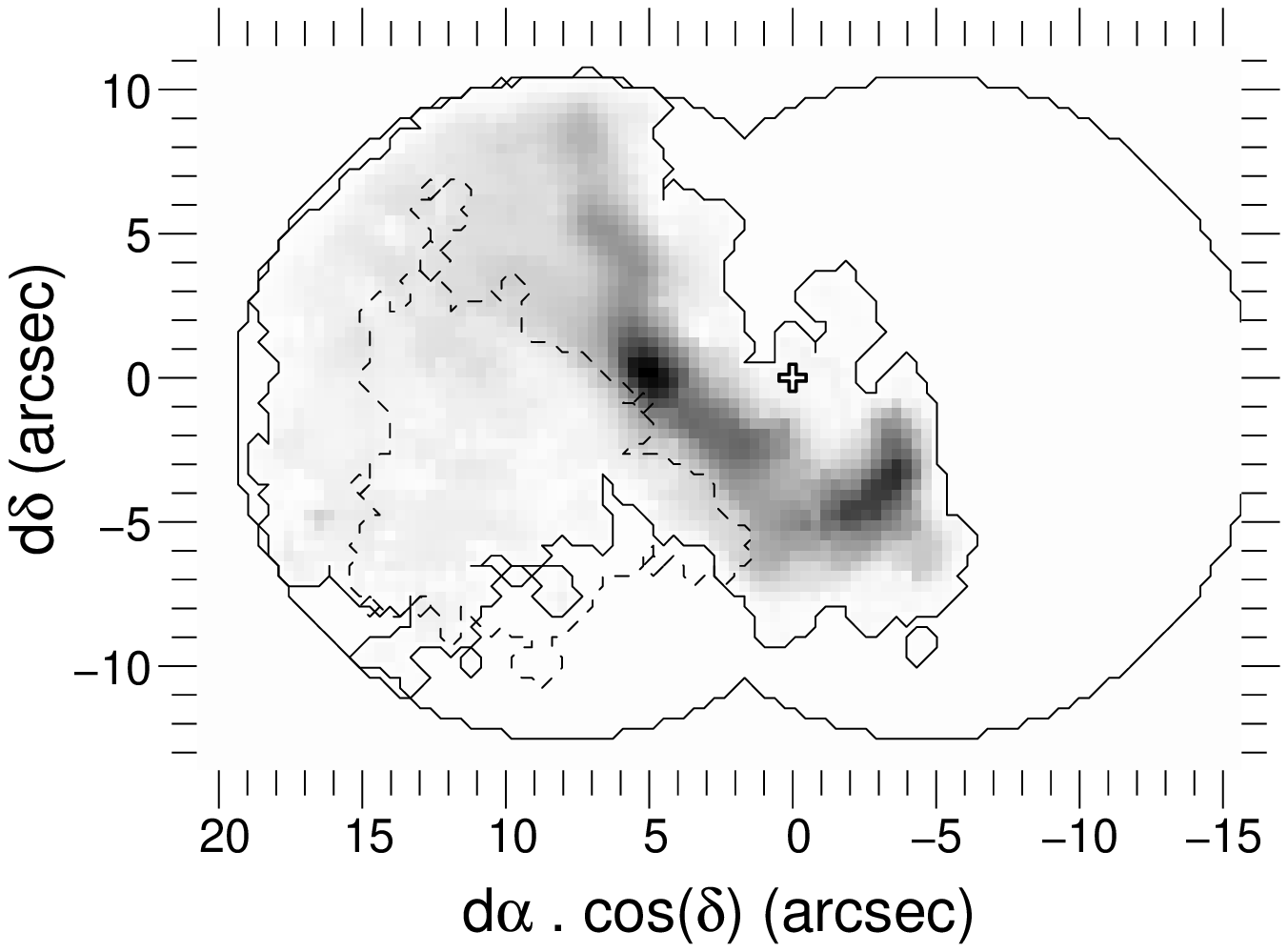}}
    \caption{(a) \object{Northern Arm}.\label{namap}\label{nafmap}\vspace{0.97cm}}
  \end{minipage}\hfill
  \resizebox{0.47\hsize}{!}{\includegraphics{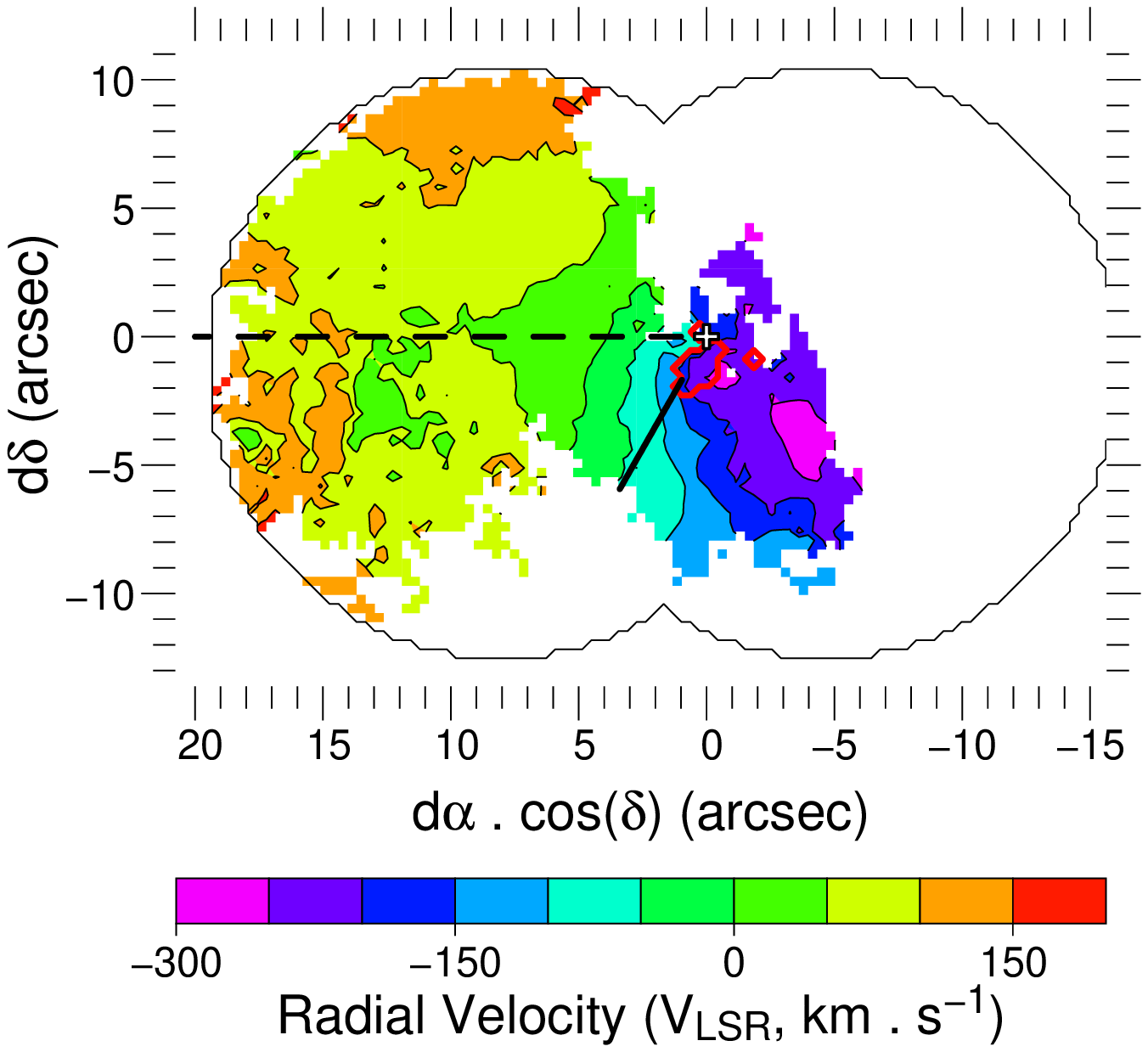}}
\end{figure*}

\begin{figure*}
  \begin{minipage}[b]{0.47\hsize}
    \resizebox{\hsize}{!}{\includegraphics{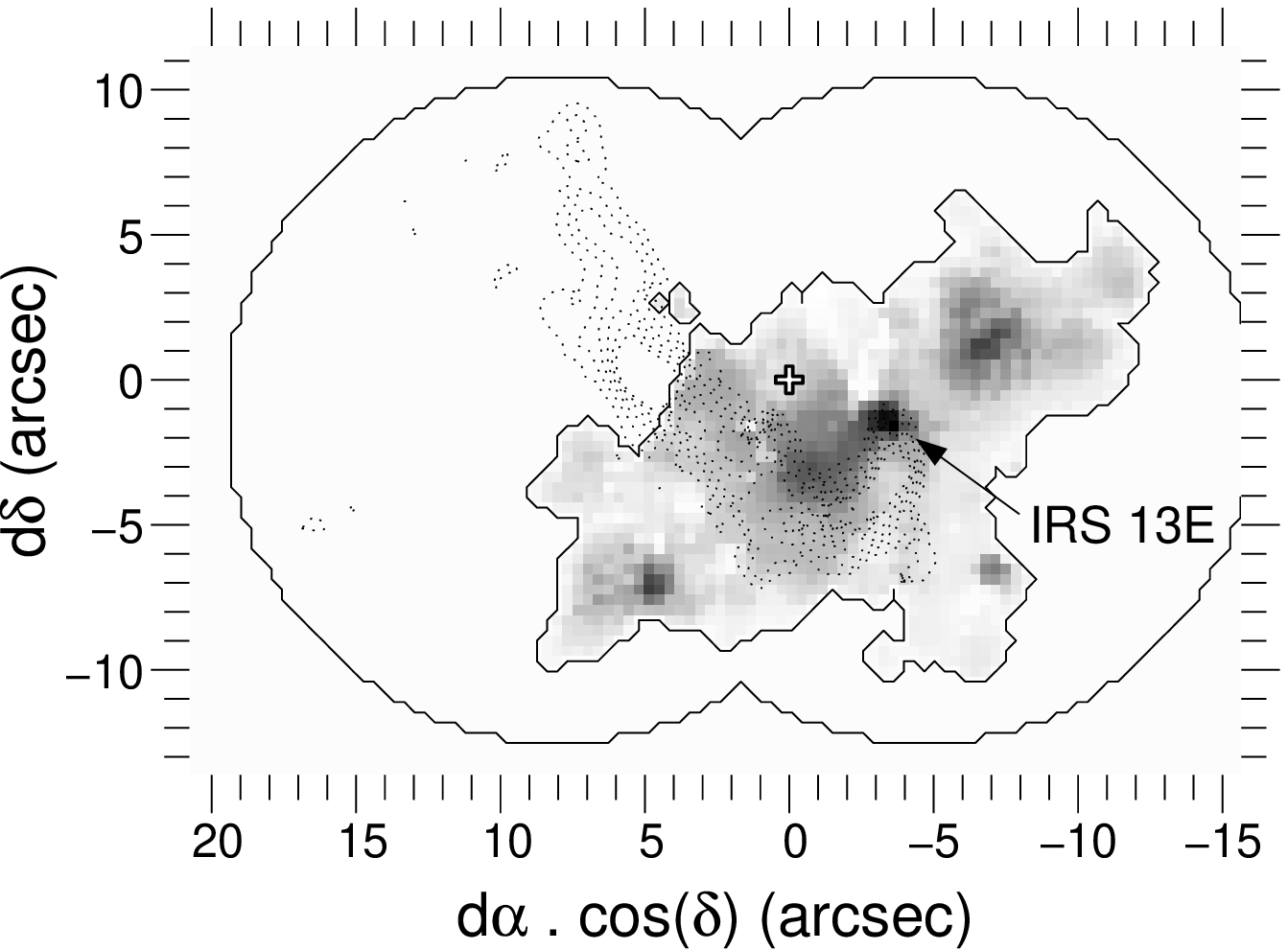}}
    \caption{(b) \object{Bar}\label{barmap}.\label{barfmap}\vspace{0.97cm}}
  \end{minipage}\hfill
  \resizebox{0.47\hsize}{!}{\includegraphics{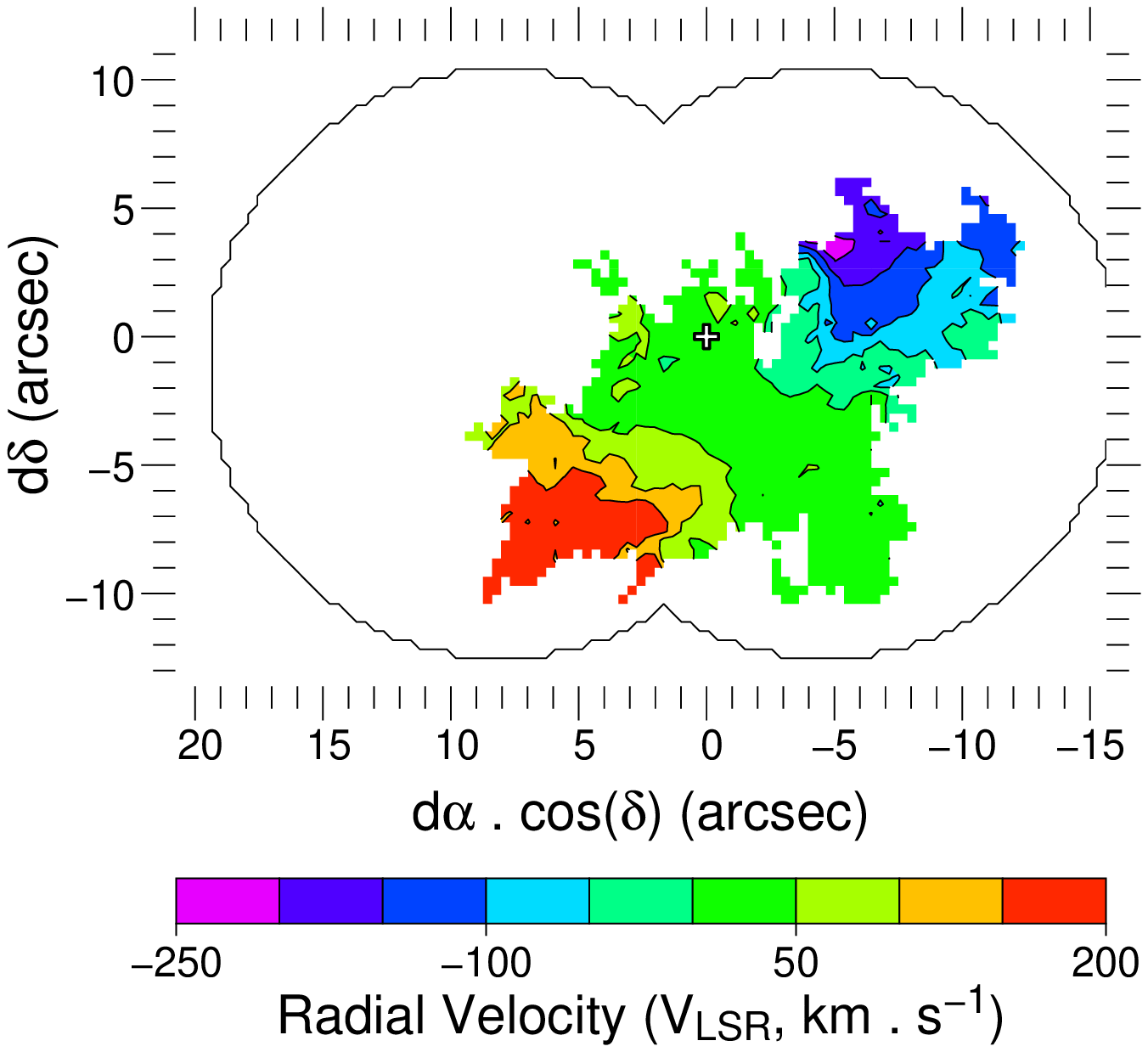}}
\end{figure*}

\begin{figure*}
  \begin{minipage}[b]{0.47\hsize}
    \resizebox{\hsize}{!}{\includegraphics{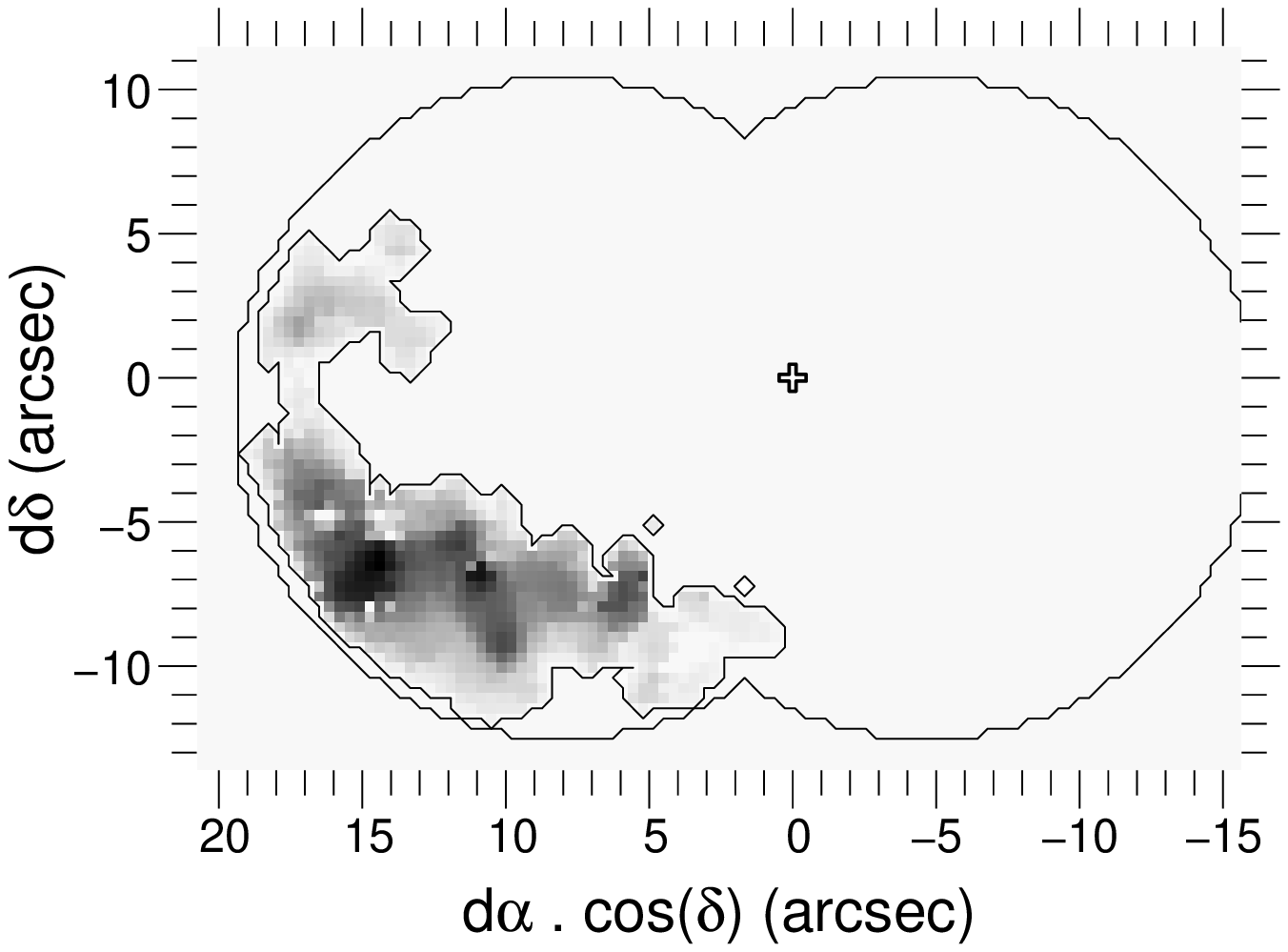}}
    \caption{(c) \object{Ribbon}.\label{eamap}\vspace{0.97cm}}
  \end{minipage}\hfill
  \resizebox{0.47\hsize}{!}{\includegraphics{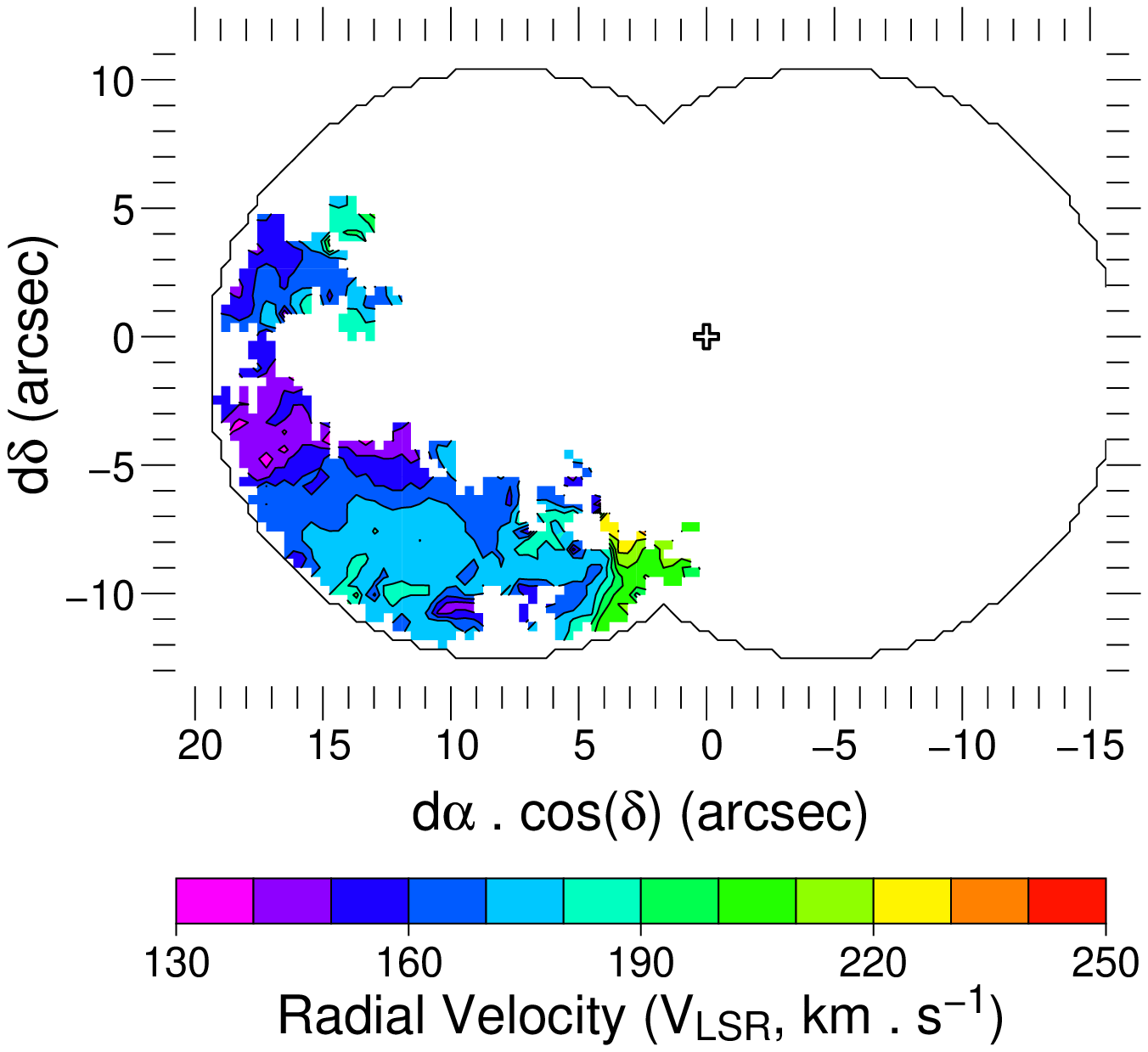}}
\end{figure*}

\begin{figure*}
  \begin{minipage}[b]{0.47\hsize}
    \resizebox{\hsize}{!}{\includegraphics{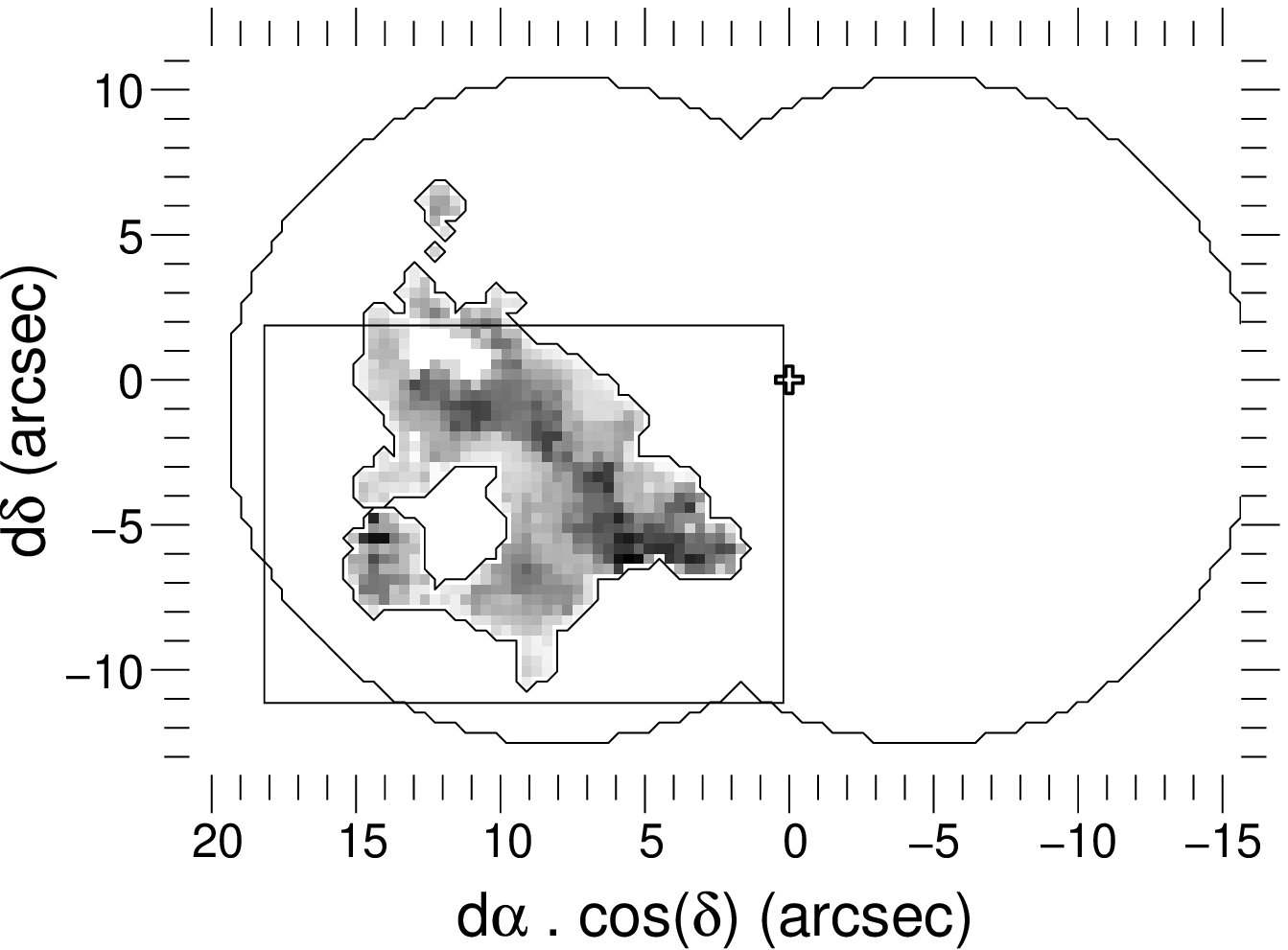}}
    \caption{(d) \object{Eastern Bridge}. The box shows the field of
      Fig.~\ref{figure:eb}.\label{ebmap}\vspace{0.97cm}}
  \end{minipage}\hfill
  \resizebox{0.47\hsize}{!}{\includegraphics{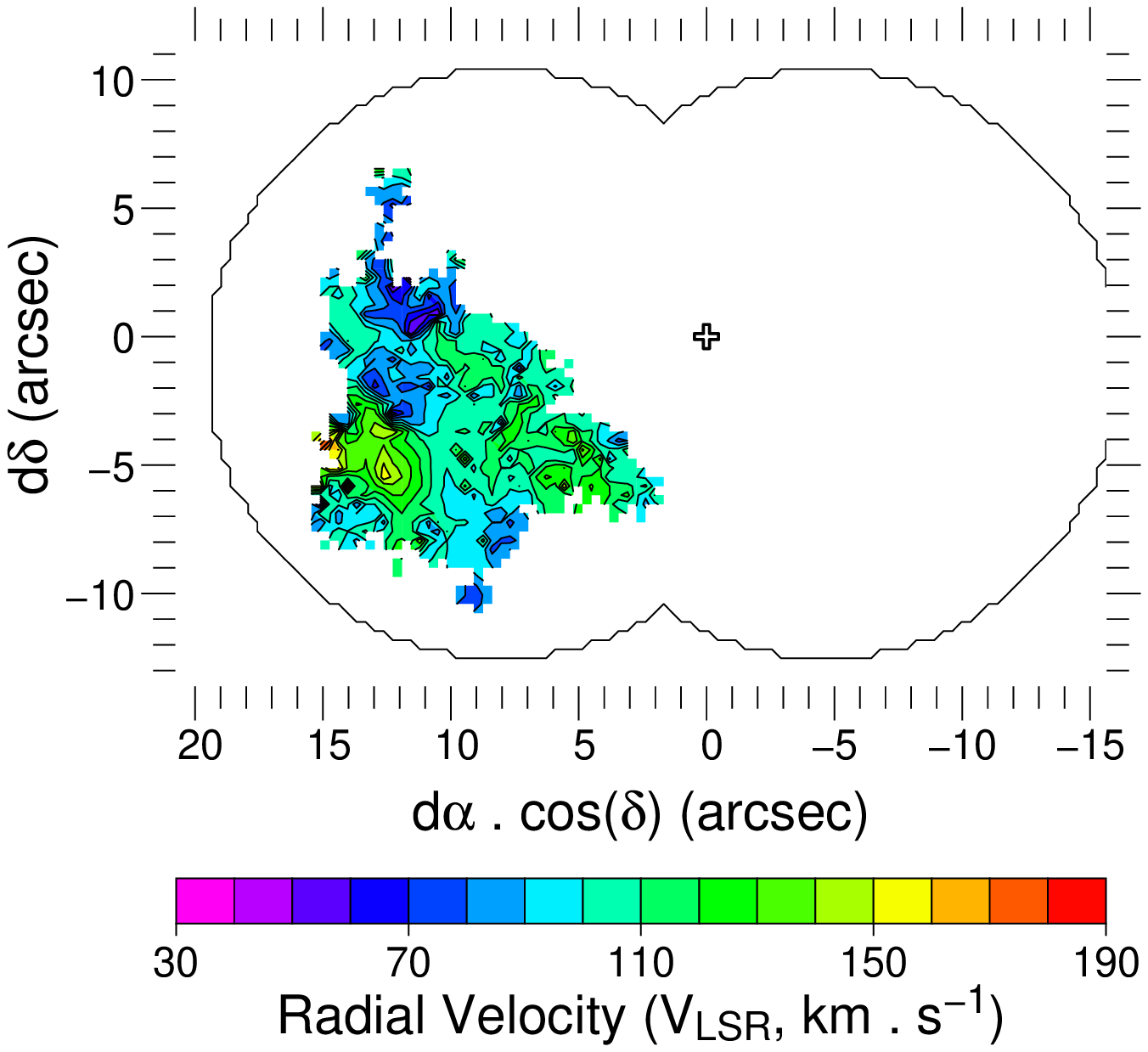}}
\end{figure*}

\begin{figure*}
  \begin{minipage}[b]{0.47\hsize}
  \resizebox{\hsize}{!}{\includegraphics{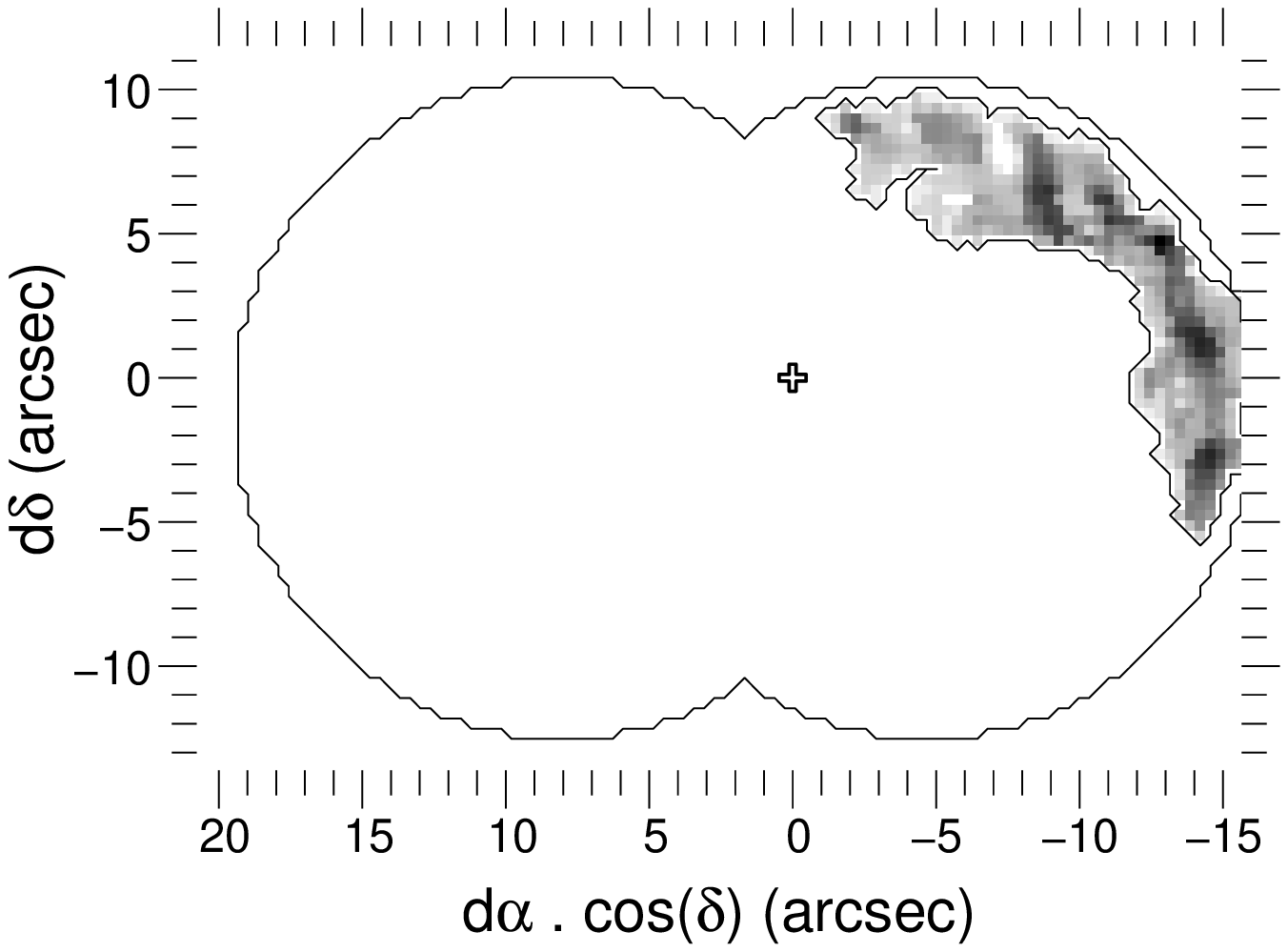}}
  \caption{(e) \object{Western Arc}.\label{bomap}\vspace{0.97cm}}
  \end{minipage}\hfill
  \resizebox{0.47\hsize}{!}{\includegraphics{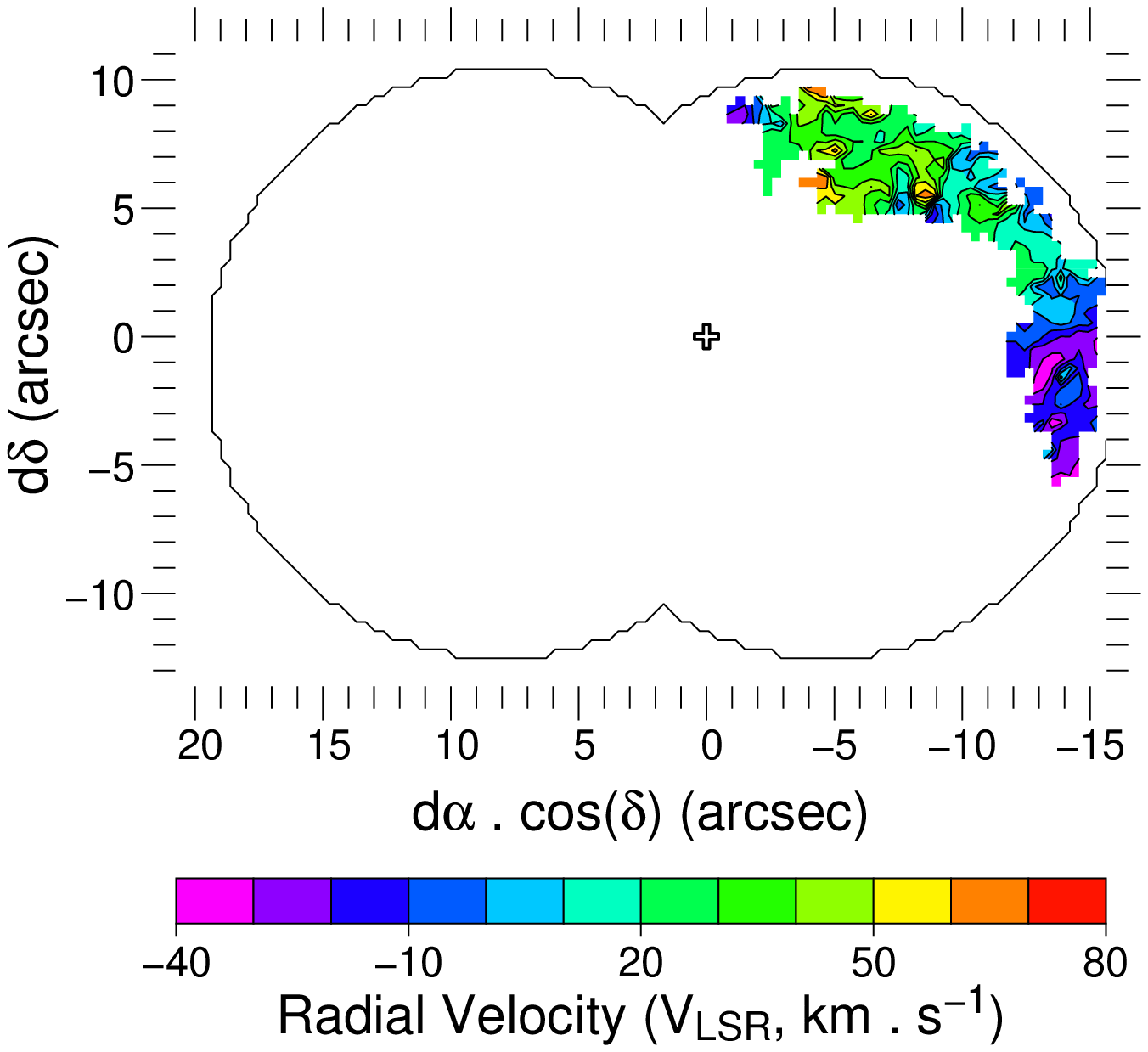}}
\end{figure*}

\begin{figure*}
  \begin{minipage}[b]{0.47\hsize}
  \resizebox{\hsize}{!}{\includegraphics{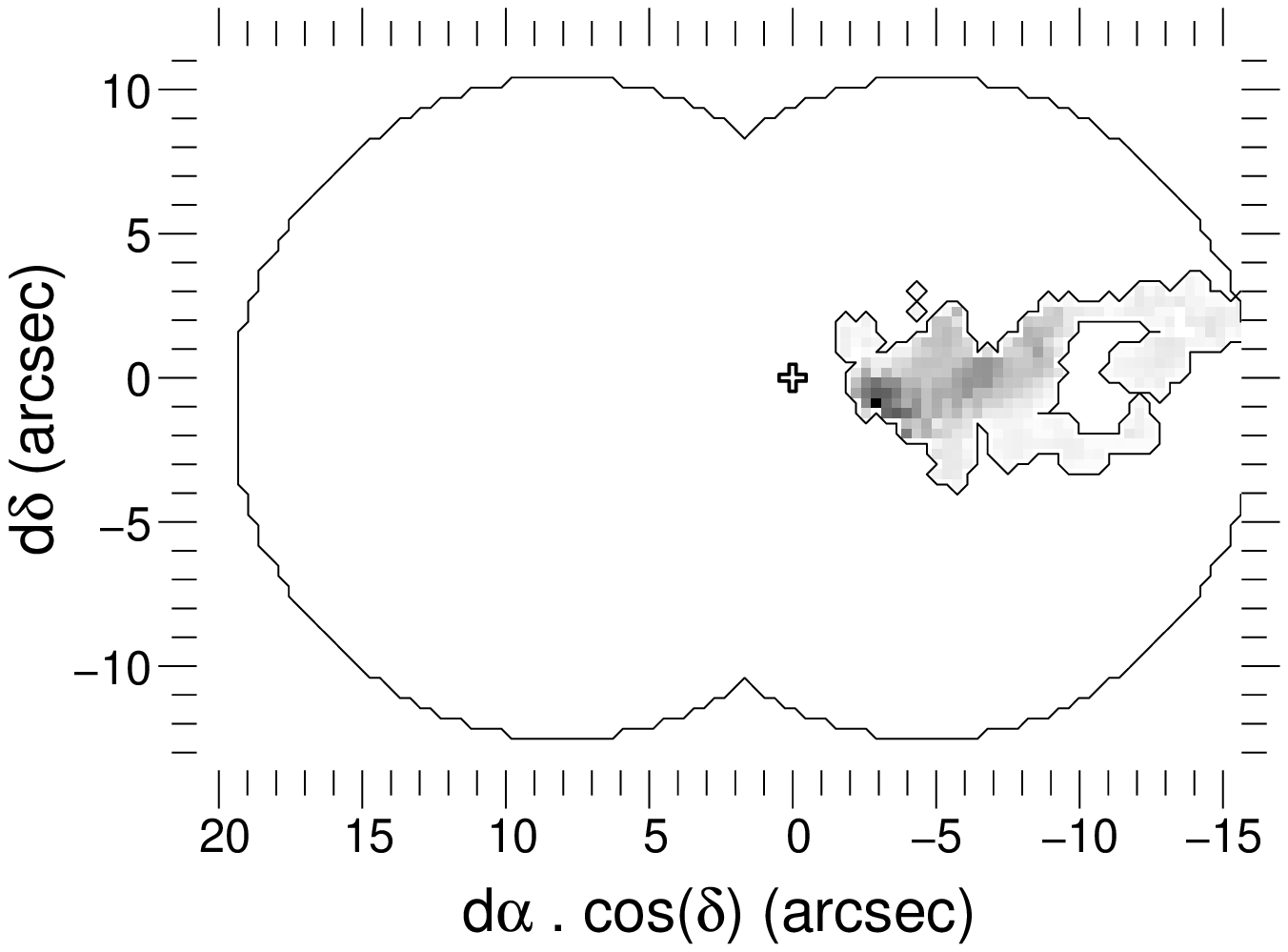}}
  \caption{(f) \object{Western Bridge}.\label{wbmap}\vspace{0.97cm}}
  \end{minipage}\hfill
  \resizebox{0.47\hsize}{!}{\includegraphics{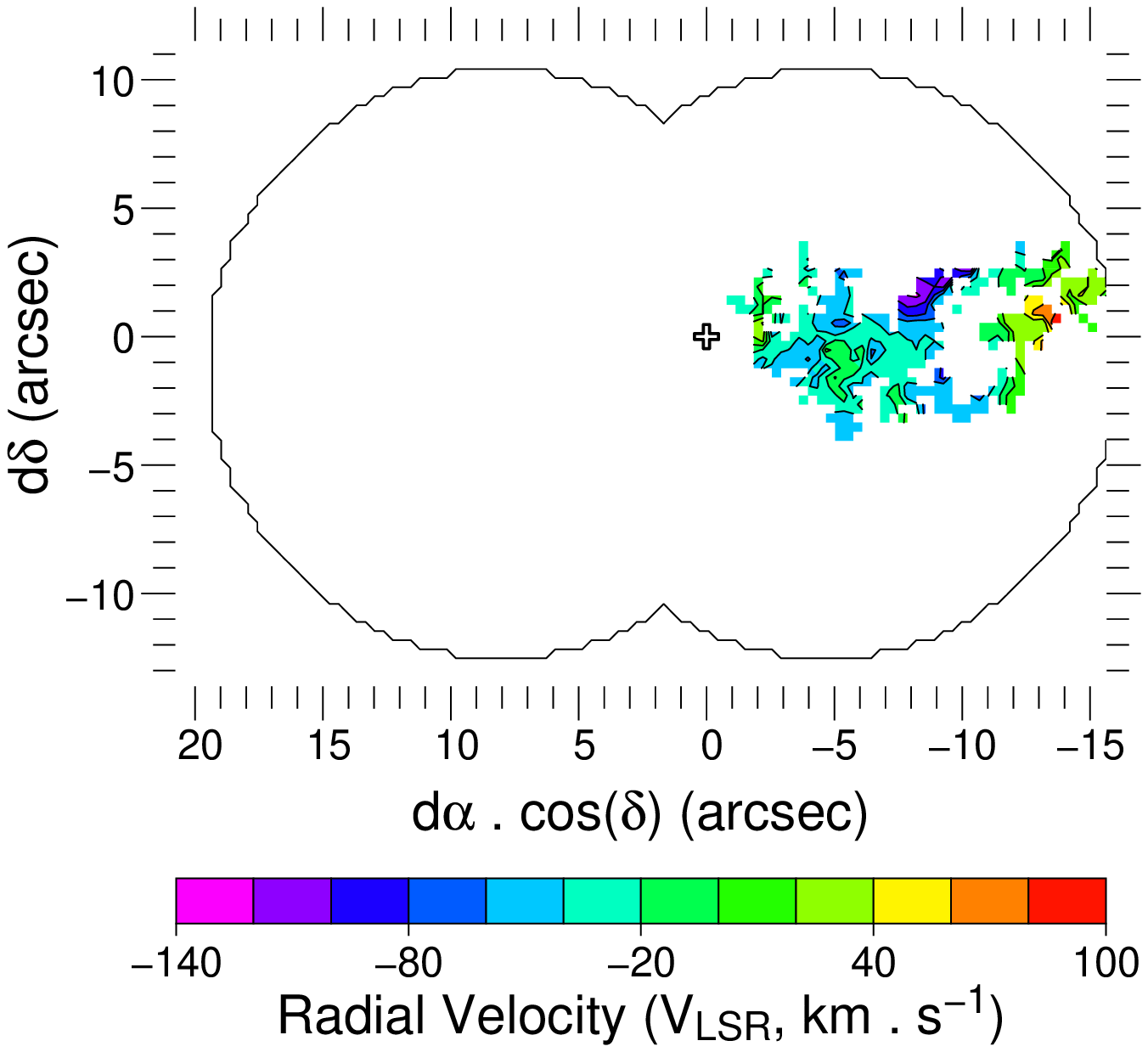}}
\end{figure*}

\begin{figure*}
  \begin{minipage}[b]{0.47\hsize}
  \resizebox{\hsize}{!}{\includegraphics{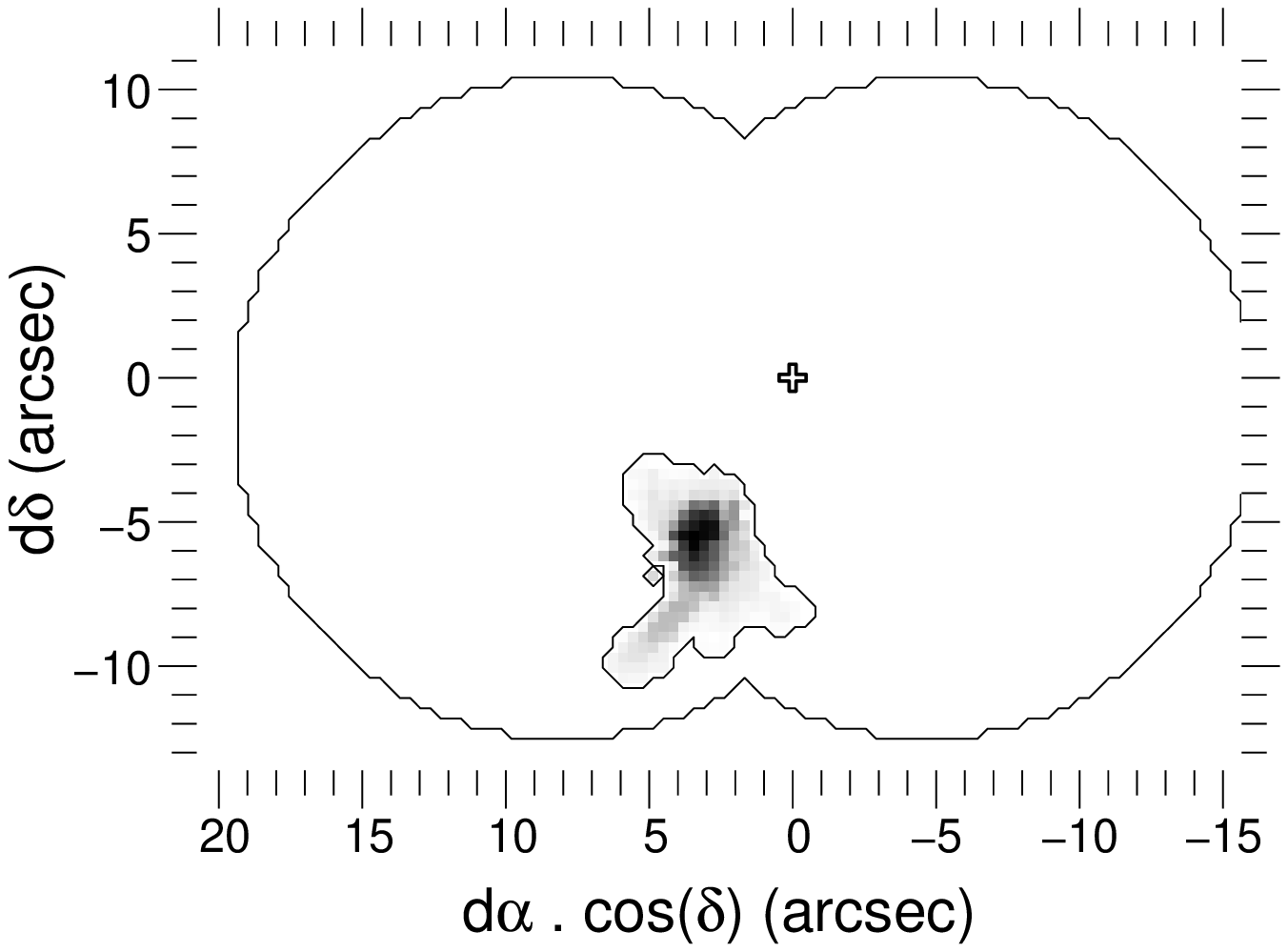}}
  \caption{(g) \object{Tip}.\label{tipmap}\vspace{0.97cm}}
  \end{minipage}\hfill
  \resizebox{0.47\hsize}{!}{\includegraphics{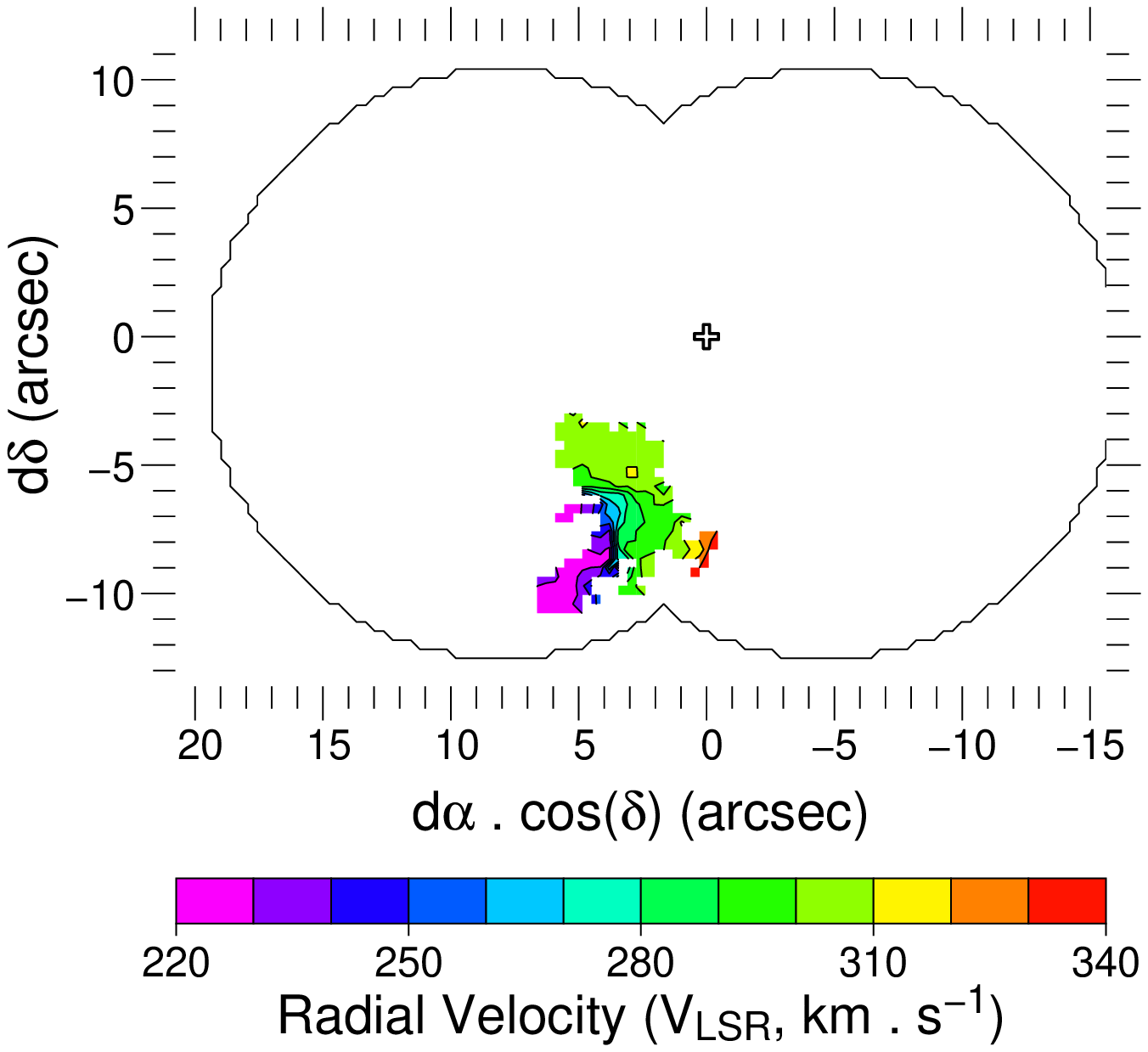}}
\end{figure*}

\begin{figure*}
  \begin{minipage}[b]{0.47\hsize}
  \resizebox{\hsize}{!}{\includegraphics{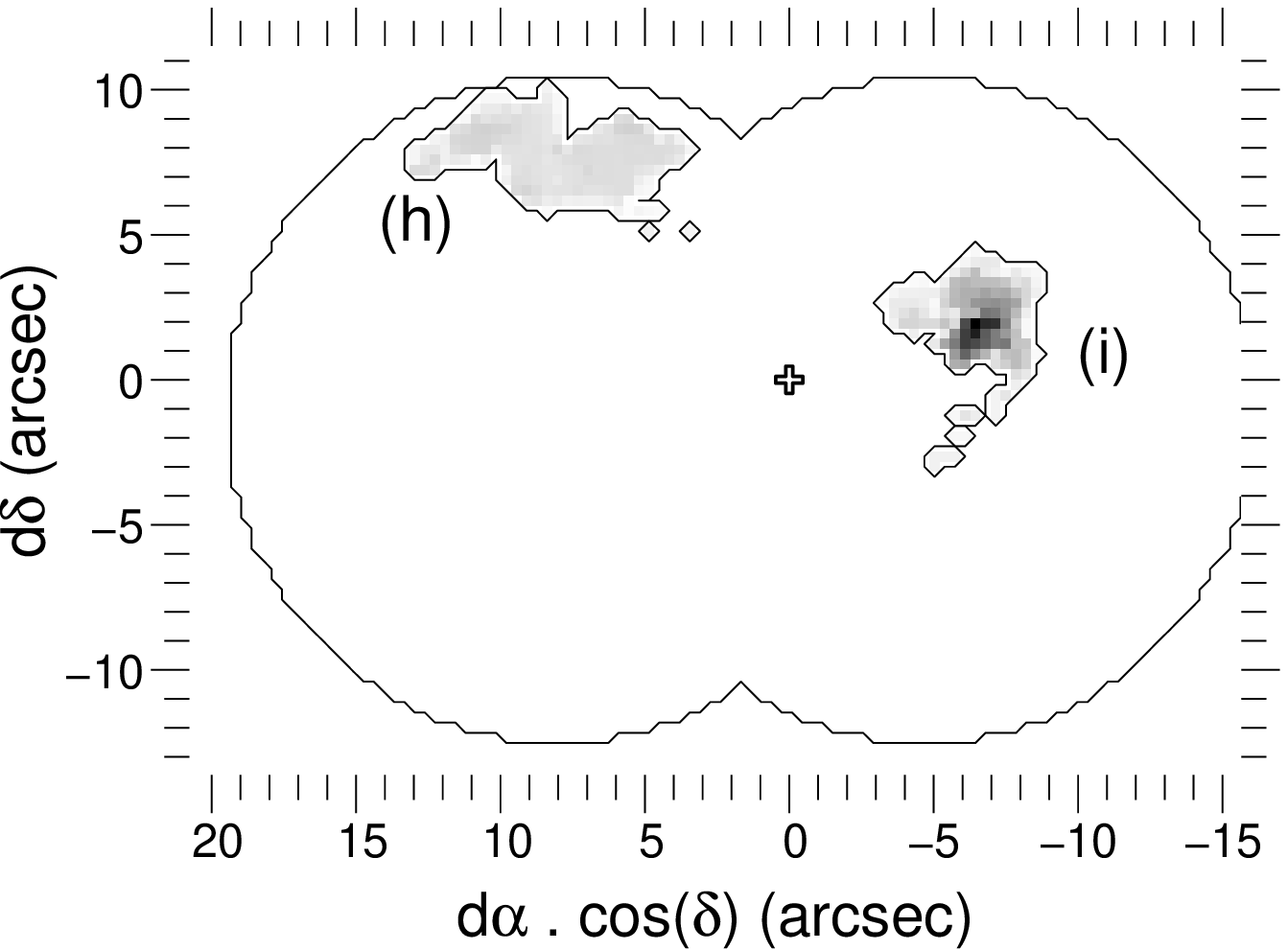}}
  \caption{(h) \object{Northern Arm Chunk} and (i) \object{Bar Overlay}.\label{nacmap}\label{bovmap}\vspace{0.97cm}}
  \end{minipage}\hfill
  \resizebox{0.47\hsize}{!}{\includegraphics{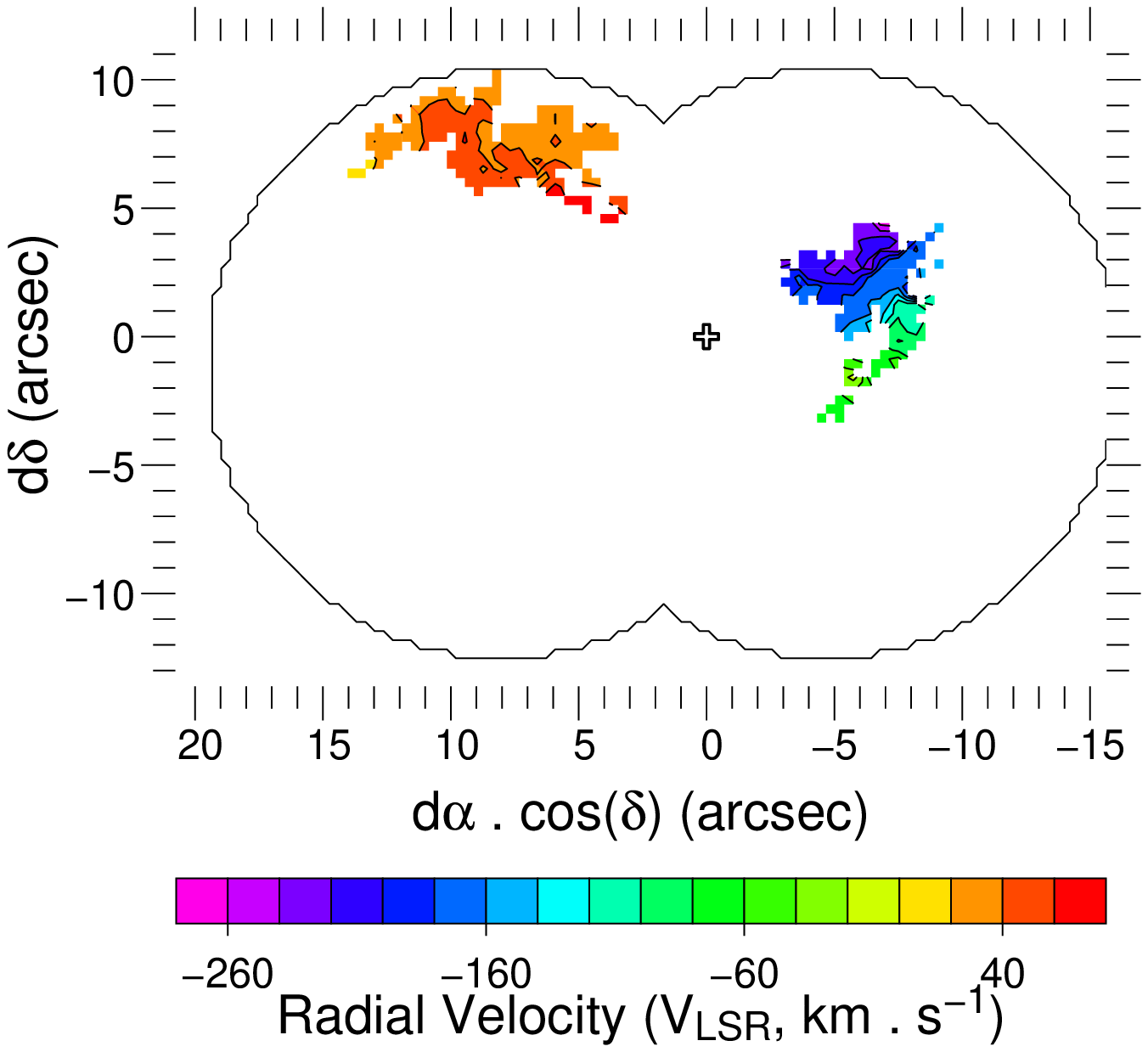}}
\end{figure*}

\section{Comparison with \ion{He}{i} data}
\label{app:heibrg}

\begin{figure}
  \includegraphics[width=\hsize]{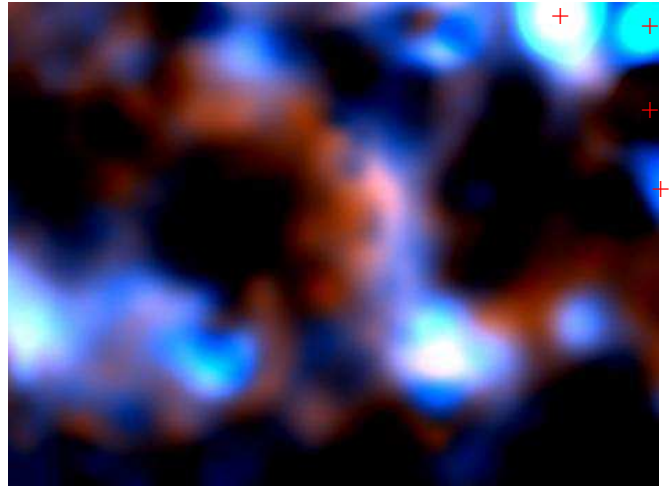}\hfill
    \caption{\label{figure:eb}Composite image of the \object{Eastern Bridge},
      from  images integrated  from 50  to 100~km~s$^{-1}$  in  the Br$\gamma$
      (red) and \ion{He}{i} (blue) cubes.   The positions of four helium stars
      at the upper right corner are  given by red crosses (\object{GCIRS~16NE}, 16C, 16SW
      and 33SE). The field is shown in Fig.~\ref{ebmap}.}
\end{figure}
%% \begin{figure*}
%%   \includegraphics[width=0.55555555\hsize]{0209fiB1.eps}\hfill
%%   \begin{minipage}[b]{0.38888888\hsize}
%%     \caption{\label{figure:eb}Composite image of the \object{Eastern Bridge},
%%       from  images integrated  from 50  to 100~km~s$^{-1}$  in  the Br$\gamma$
%%       (red) and \ion{He}{i} (blue) cubes.   The positions of four helium stars
%%       at the upper right corner are  given by red crosses (\object{GCIRS~16NE}, 16C, 16SW
%%       and 33SE). The field is shown in Fig.~\ref{ebmap}.}
%%   \end{minipage}
%% \end{figure*}

\textsf{(a)} Though the {Northern Arm} remains the most prominent feature of the
\object{Minispiral}, the mean value of its normalized [\ion{He}{i}]/[Br$\gamma$]
line ratio (Minicavity excluded) is one of the lowest ($\simeq0.74$), being only
higher than  the value measured for  the small part  of the Western Arc  that we
detect.   Considering  that the  faintest  parts of  the  Northern  Arm are  not
detected in  \ion{He}{i}, this  value may  be even smaller.   The line  ratio is
higher on the  western side of the  bright rim, and in \ion{He}{i}  this rim has
the shape of a part of a circle surrounding the \object{GCIRS~16} cluster.  This
circle continues  further to the northwest,  forming a rather faint  horn at the
location where, in  Br$\gamma$, the rim bends abruptly  ($5\arcsec$ to the north
and $5\arcsec$ to the east of Sgr~A$^\star$, spot C
on Fig.~\ref{PaumardT-fig5}).\\
The open ring  of ionized gas surrounding the Minicavity  is on average brighter
in \ion{He}{i}  than the rest  of the {Northern  Arm} relative to  the intensity
distribution in Br$\gamma$. Its innermost  border is even brighter.  Its western
edge,  where \object{GCIRS~13}  and \object{GCIRS~2}  lie, is  very  bright, and
looks like a vertical bar going from \object{GCIRS~13} almost to the declination
of  the  \object{AF  star},   making  the  Minicavity  look  angular.   Physical
implications  of   these  variations  of   the  line  ratio  are   discussed  in
Sect.~\ref{discussion}.

\textsf{(b)}    The   Bar    is    the   main    feature    with   the    highest
  [\ion{He}{i}]/[Br$\gamma$] ratio, with a normalized value of 0.99.  However,
  we  do  not  detect  helium  towards  the  full  extent  of  its  Br$\gamma$
  counterpart.

\textsf{(d)} The Eastern Bridge (Fig.~\ref{figure:eb}) is clearly identified, but
  it presents a  shape much different from the one  observed in Br$\gamma$. It
  is brighter on its southern side, and the northern parts are not detected by
  the procedure.   The southern parts extend horizontally,  following the edge
  of the  Eastern Arm Ribbon  upon which it  is superimposed, with  a velocity
  offset between  the Eastern Bridge  and the Ribbon of  about -50~km~s$^{-1}$
  (measured in Br$\gamma$,  but the agreement is good  between the two lines),
  which  again suggests  that the  two features  are related.   The bow-shaped
  bright rim of the structure, which  is almost vertical and gives its name to
  the Eastern  Bridge, is  offset in  \ion{He}{i} by about  $1\arcsec$ to  the West
  relative to Br$\gamma$.

\textsf{(e)} A  small part of the Western  Arc is detected within  our field; its
  [\ion{He}{i}]/[Br$\gamma$] value is the smallest,  but only a few points are
  detected both in \ion{He}{i} and Br$\gamma$.

\textsf{(g)}  Due  to the  lower  spectral  resolution  of the  \ion{He}{i}  data
  (52.9~km~s$^{-1}$,  vs.  21.3~km~s$^{-1}$  in  Br$\gamma$), the  Tip is  not
  separated from  the Ribbon by  our procedure in  this band.  However,  it is
  clearly  seen.  It  is  the  brightest feature  relative  to its  Br$\gamma$
  counterpart, with a  normalized line ratio of $\simeq2.64$.   The line ratio
  is  also  noticeably   brighter  on  its  southwestern  edge   than  on  its
  northeastern edge, which is not detected in \ion{He}{i} by the decomposition
  procedure.  The \object{Microcavity} is also observed in \ion{He}{i}.

\end{document}